\documentclass[twocolumn,trackchanges]{aastex631}

\usepackage{CJK}
\newcommand{\petit}{\texttt{petitRADTRANS}}

\newcommand{\dynesty}{\texttt{dynesty}}
\newcommand{\teff}{T$_{\rm eff}$}

\newcommand{\kms}{$\rm km\,s^{-1}$}
\newcommand{\HDb}{HD~143105~b}
\newcommand{\caltech}{Department of Astronomy, California Institute of Technology, Pasadena, CA 91125, USA}
\newcommand{\gps}{Division of Geological \& Planetary Sciences, California Institute of Technology, Pasadena, CA 91125, USA}
\newcommand{\ucsc}{Department of Astronomy \& Astrophysics, University of California, Santa Cruz, CA95064, USA}
\newcommand{\keck}{W. M. Keck Observatory, 65-1120 Mamalahoa Hwy, Kamuela, HI 96743, USA}
\newcommand{\ucla}{Department of Physics \& Astronomy, 430 Portola Plaza, University of California, Los Angeles, CA 90095, USA}
\newcommand{\jpl}{Jet Propulsion Laboratory, California Institute of Technology, 4800 Oak Grove Dr.,Pasadena, CA 91109, USA}
\newcommand{\ucsd}{Center for Astrophysics and Space Sciences, University of California, San Diego, La Jolla, CA 92093}

\newcommand{\northwestern}{Center for Interdisciplinary Exploration and Research in Astrophysics (CIERA) and Department of Physics and Astronomy,
Northwestern University, Evanston, IL 60208, USA}
\newcommand{\arizona}{James C. Wyant College of Optical Sciences, University of Arizona,
Meinel Building 1630 E. University Blvd., Tucson, AZ 85721, USA}

\received{\today}
\shorttitle{Inclination and composition of HD~143105~b}
\shortauthors{Finnerty et al.}
\graphicspath{{./}{}}
\begin{document}
\begin{CJK*}{UTF8}{gbsn}

\title{True mass and atmospheric composition of the non-transiting hot Jupiter HD 143105 b}

\correspondingauthor{Luke Finnerty}
\email{lfinnerty@astro.ucla.edu}

\author[0000-0002-1392-0768]{Luke Finnerty}
\affiliation{\ucla}

% Xin
\author[0000-0002-6171-9081]{Yinzi Xin}
\affiliation{\caltech}

% Xuan
\author[0000-0002-6618-1137]{Jerry W. Xuan}
\affiliation{\caltech}

% Inglis
\author{Julie Inglis}
\affiliation{\caltech}

% Fitzgerald
\author[0000-0002-0176-8973]{Michael P. Fitzgerald}
\affiliation{\ucla}

\author[0000-0003-2429-5811]{Shubh Agrawal}
\affiliation{Department of Physics and Astronomy, University of Pennsylvania, Philadelphia, PA 19104, USA}

\author[0000-0002-6525-7013]{Ashley Baker}
\affiliation{\caltech}

\author{Randall Bartos}
\affiliation{\jpl}

\author{Geoffrey A. Blake}
\affiliation{\gps}

% Bond
% \author{Charlotte Z. Bond}
% \affiliation{UK Astronomy Technology Centre, Royal Observatory, Edinburgh EH9 3HJ, United Kingdom}

% Calvin
\author[0000-0003-4737-5486]{Benjamin Calvin}
\affiliation{\caltech}
\affiliation{\ucla}

% Cetre
\author{Sylvain Cetre}
\affiliation{\keck}

% Delorme
\author[0000-0001-8953-1008]{Jacques-Robert Delorme}
\affiliation{\keck}
\affiliation{\caltech}

% Doppman
\author{Greg Doppmann}
\affiliation{\keck}

\author[0000-0002-1583-2040]{Daniel Echeverri}
\affiliation{\caltech}

\author[0000-0001-9708-8667]{Katelyn Horstman}
\affiliation{\caltech}
\altaffiliation{NSF Graduate Research Fellow}

\author[0000-0002-5370-7494]{Chih-Chun Hsu}
\affiliation{\northwestern}

% Jovanovic
\author[0000-0001-5213-6207]{Nemanja Jovanovic}
\affiliation{\caltech}

% Liberman
\author[0000-0002-4934-3042]{Joshua Liberman}
\affiliation{\caltech}
\affiliation{\arizona}

% Lopez
\author[0000-0002-2019-4995]{Ronald A. L\'opez}
\affiliation{\ucla}

% Martin
\author[0000-0002-0618-5128]{Emily C. Martin}
\affiliation{\ucsc}

% Mawet
\author{Dimitri Mawet}
\affiliation{\caltech}
\affiliation{\jpl}

% Morris
\author{Evan Morris}
\affiliation{\ucsc}

% Pezzato
\author{Jacklyn Pezzato-Rovner}
\affiliation{\caltech}

% Ruffio
\author[0000-0003-2233-4821]{Jean-Baptiste Ruffio}
\affiliation{\ucsd}

% Sappey
\author[0000-0003-1399-3593]{Ben Sappey}
\affiliation{\ucsd}

% Schofield
\author{Tobias Schofield}
\affiliation{\caltech}

% Skemer
\author{Andrew Skemer}
\affiliation{\ucsc}

% Venenciano
\author{Taylor Venenciano}
\affiliation{Physics and Astronomy Department, Pomona College, 333 N. College Way, Claremont, CA 91711, USA}

% Wallace
\author[0000-0001-5299-6899]{J. Kent Wallace}
\affiliation{\jpl}

% Wallack
\author[0000-0003-0354-0187]{Nicole L. Wallack}
\affiliation{Earth and Planets Laboratory, Carnegie Institution for Science, Washington, DC 20015, USA}

% Wang
\author[0000-0003-0774-6502]{Jason J. Wang (王劲飞)}
\affiliation{\northwestern}

% Wang (王吉)
\author[0000-0002-4361-8885]{Ji Wang (王吉)}
\affiliation{Department of Astronomy, The Ohio State University, 100 W 18th Ave, Columbus, OH 43210 USA}



\begin{abstract}
We present Keck/KPIC phase II $K$-band observations of the non-transiting hot Jupiter \HDb. Using a cross-correlation approach, we make the first detection of the planetary atmosphere at $K_p = 185^{+11}_{-13}$ \kms\ and an inferior conjunction time 2.5 hours before the previously-published ephemeris. The retrieved $K_p$ value, in combination with orbital period, mass of the host star, and lack of transit detection, gives an orbital inclination of $78^{\circ+2}_{-12}$ and a true planet mass of 1.23$\pm0.10\rm\ M_J$. While the equilibrium temperature of \HDb\ is in the transition regime between non-inverted and inverted atmospheres, our analysis strongly prefers a non-inverted atmosphere. Retrieval analysis indicates the atmosphere of \HDb\ is cloud-free to approximately 1 bar and dominated by H$_2$O absorption ($\log \rm H_2O_{MMR} = -3.9^{+0.8}_{-0.5}$), placing only an upper limit on the CO abundance ($\log \rm CO_{MMR} < -3.7$ at 95\% confidence). We place no constraints on the abundances of Fe, Mg, or $^{13}$CO. From these abundances, we place an upper limit on the carbon-to-oxygen ratio for \HDb, $\rm C/O < 0.2$ at 95\% confidence, and find the atmospheric metallicity is approximately $0.1\times$ solar. The low metallicity may be responsible for the lack of a thermal inversion, which at the temperature of \HDb\ would likely require significant opacity from TiO and/or VO. With these results, \HDb\ joins the small number of non-transiting hot Jupiters with detected atmospheres.

\end{abstract}



\section{Introduction} \label{sec:intro}
Transmission and secondary eclipse spectroscopy are powerful tools for characterizing exoplanet atmospheres, but are fundamentally limited to the small fraction of planets which transit their host stars. By directly detecting planetary dayside emission, high-resolution cross-correlation spectroscopy (HRCCS) techniques can be used to characterize the atmospheres of non-transiting exoplanets \citep[e.g.][]{brogi2012, rodler2012, brogi2014, birkby2017, guilluy2019, webb2020, buzard2020, pelletier2021}. These observations break the mass-inclination degeneracy from the  radial velocity technique and enable constraints on the true mass of non-transiting planets. The significant inclinations of some of these systems may also enable novel characterization of the polar regions of hot Jupiters \citep{malsky2021}. 

High-resolution spectroscopy of non-transiting planets must contend with additional uncertainties compared with observations of transiting targets, increasing the risks associated with these observations. For planets with nearly face-on orbits, the projected line-of-sight velocity shift may be too small compared to the instrument resolution to make a detection, even if the planet would be detectable if the orbit were edge on. The observed velocity shift decreases with $\sin i$, so this effect is prohibitive for a relatively small fraction of non-transiting planets, but necessitates additional care when scheduling observations so that a sufficient velocity shift will be obtained even if the system is low inclination.

Additionally, non-transiting planets have ephemerides based on radial velocity observations alone. While transiting planets detected with TESS often have periods and transit times determined to a precision of seconds, such precise ephemerides are difficult to achieve for non-transiting planets. Furthermore, maintaining current ephemerides via radial velocity monitoring requires significant investments of telescope time, and as a result many non-transiting systems do not have recently updated ephemerides. This can lead to significant uncertainty in the orbital phase for followup HRCCS observations.

These challenges have limited the number of non-transiting planets detected via HRCCS. $\tau$~Boo~b was the first non-transiting planet detected with HRCCS by \citet{brogi2012}, followed by HD~179949~b \citep{brogi2014}. \citet{brogi2013} reported a tentative detection of 51~Peg~b which was confirmed by \citet{birkby2017}, and HD~102195~b was first reported in \citet{guilluy2019}. HD~187123~b was detected via a similar technique by \citet{buzard2020}. This is only a small fraction of the number of transiting systems characterized with HRCCS techniques.

\HDb\ is a non-transiting hot Jupiter first detected in 2016 using radial velocity observations from the SOPHIE spectrograph \citep{hebrard2016}. Properties of the system are summarized in Table \ref{tab:props}. While the short orbital period and bright host star make \HDb\ a good candidate for atmospheric characterization, the far-northern declination of the system has made it inaccessible to most ground-based high-resolution spectrographs, and \textit{JWST} has so far only characterized hot Jupiters using transmission and eclipse methods, leaving \HDb\ unstudied. \HDb\ is a particularly interesting target for atmospheric characterization, as its 1940 K equilibrium temperature is close to the expected transition between atmospheres with and without a thermal inversion predicted by both 1D \citep{baxter2020} and 3D modeling \citep{roth2024}. In this paper, we use Keck/KPIC HRCCS observations to make the first atmospheric detection of \HDb\, which enables us to refine the planetary ephemeris and constrain the atmospheric composition and pressure-temperature ($P-T$) profile. 

We discuss the Keck/KPIC observations of \HDb\ and our atmospheric retrieval framework in Section \ref{sec:obs}. Results from the cross-correlation and retrieval analyses are presented in Section \ref{sec:res}, and discussed in Section \ref{sec:disc}. Section \ref{sec:conc} concludes.  

 \begin{deluxetable}{ccc}
    \tablehead{\colhead{Property} & \colhead{Value} & \colhead{Ref.}}
    \startdata
        & \textbf{HD 143105} & \\
        \hline
        RA & 15:53:36.57 &  \citet{gaiaedr3} \\
        Dec & +68:43:11.92 &  \citet{gaiaedr3} \\
        Sp. Type & F5 & \citet{hebrard2016} \\
        $K_{\rm mag}$ & $5.52\pm0.02$ & \citet{cutri2003}  \\
        Mass & $1.51\pm0.11\ M_\odot$ & \citet{hebrard2016}   \\
        Radius & $1.61^{+0.08}_{-0.05} \rm\ R_\odot$ & \citet{exofop3} \\
        \teff & $6380\pm60$ K & \citet{hebrard2016} \\
        $\log g$ & $4.37\pm0.04$ & \citet{hebrard2016}   \\
        $v\sin i$ & $9.1\pm1$\ \kms & \citet{hebrard2016} \\
        $v_{\rm rad}$ & $15.942\pm0.003$ \kms & \citet{hebrard2016}  \\
        $\rm [Fe/H]$ & $0.05\pm0.04$ & \citet{hypatia}  \\
        $\rm [C/H]$ & $0.03\pm0.34$ & \citet{hypatia} \\
        $\rm [O/H]$ & $0.02\pm0.04$ & \citet{hypatia} \\
        % C/O &  & \\
        \smallskip \\
        \hline
         & \textbf{HD 143105 b} & \\
        \hline
        Period &  $2.1974\pm0.0003$ days & \citet{hebrard2016} \\
        $\rm T_{\rm conj}$ & JD $2456531.344\pm0.007$ & \citet{hebrard2016}  \\
        $\rm T_{\rm conj}$ & JD $2456531.237^{+0.012}_{-0.017}$ & This work \\
        $a$ & $0.0379\pm0.0009$ AU &  \citet{hebrard2016}  \\
        $e$ & $<0.07$ & \citet{hebrard2016} \\
        $i$ & $78^{\circ +2}_{-12}$ & This work \\
        $K_{\rm p}$ & $185^{+11}_{-13}$ \kms & This work  \\
        $\rm M\sin i$ & $1.21\pm0.06 \rm\ M_J$  & \citet{hebrard2016} \\
        Mass & $1.23\pm0.10 \rm\ M_J$ & This work \\
        Radius & $1.20\pm0.05$\ $R_J$ & Assumed \\
        $\rm T_{\rm eq}$ & 1940 K & Calculated  \\
        C/O & $<0.4$ & This work  \\
        $\rm [C/H]$ & $<-1.1$ &  This work \\
        $\rm [O/H]$ & $-1.3^{+0.9}_{-0.5}$ &  This work \\
        $v\sin i$ &  $7.4^{+3.1}_{-3.5}$ \kms & This work
    \enddata
    \caption{Stellar and planetary properties for the HD 143105 system. Upper limits from this work are given at 95\% confidence, omitting the secondary mode for the C/O ratio.} 
    \label{tab:props}
\end{deluxetable}

\section{Observations and Data Reduction}\label{sec:obs}

\subsection{Observations}

The HD~143105 system was observed using Keck II/KPIC phase II \citep{nirspec, nirspecupgrade, nirspecupgrade2, kpic, echeverri2022} for the first half of the night on UT 24 May 2024, from 6:24 to 10:19. The observations were scheduled around a predicted superior conjunction based on the \citet{hebrard2016} ephemeris in order to maximize dayside visibility and velocity shift of the planet over the observation. The maximum velocity shift was estimated to be 80 \kms\ for a 90 degree inclination, sufficient to allow detection of the planet even if a significant orbital inclination resulted in a much smaller change in the projected velocity over the course of the observation.  Exposures were taken staring on KPIC science fiber 2 with a 90 second integration time in order to avoid saturation on the NIRSPEC detector. 

Weather conditions were sub-optimal throughout the observation. Telescope opening was delayed by approximately 40 minutes at the start of the night due to high humidity, which persisted throughout the night. Cloud cover was variable throughout the observations, which resulted in brief periods of extinction up to two magnitudes. The observation sequence began as HD~143105 was rising, at an airmass of 1.93, and continued until the target was just past transit, at an airmass of 1.51. Despite the high airmass, the top-of-atmosphere throughput to KPIC was consistently $\sim2.5$\% during clear periods, consistent with the median KPIC phase II performance \citep[][Jovanovic et al. in prep.]{echeverri2022}. This is a significant improvement in instrument performance over previous high-airmass observations taken under better atmospheric conditions. This improvement is likely as a result of recent upgrades to the real-time controller for the Keck AO system providing better AO and differential refraction corrections at high airmass. 

\subsection{Data Reduction}
Similar to previous KPIC observations \citep[e.g.][]{finnerty2023, finnerty2024}, we observed a late-type star to provide additional stellar lines for wavelength calibration, in this case HIP 62944. The 1D spectrum was extracted and used to fit a wavelength solution following the standard KPIC Data Reduction Pipeline (DRP)\footnote{\href{https://github.com/kpicteam/kpic_pipeline/}{https://github.com/kpicteam/kpic\_pipeline/}}.

For the observations of HD~143105, we used the LSF fitting procedure described in Finnerty et al. (subm.), which handles the non-Gaussian wings and wavelength-dependence of the KPIC trace using a Gaussian-Hermite model to produce weights for optimal extraction. For the LSF, we scale the overall width of the resulting spatial profile by a factor of 1.14, as was previously done in Finnerty et al. (subm.), based on the spatial/spectral width asymmetry reported in \citet{Finnerty2022}. This LSF kernel is then convolved with the forward model of the planetary spectrum after the rotational broadening kernel is applied and prior to the calculation of the log likelihood. In theory, this technique should enable constraints on $v\sin i$ better than the 9 \kms\ resolution of the Keck/NIRSPEC spectrograph, although in practice systematic uncertainties in the LSF make physical interpretation of such small rotational velocities suspect.

Following extraction, we coadded each consecutive pair of science frames in order to boost signal-to-noise and reduce the data volume for retrieval. The final extracted SNR per wavelength channel was approximately 265 in the bluest order included in analysis and 166 in the reddest order. The decreased SNR at longer wavelengths is due to a combination of atmospheric dispersion and increased thermal background at the red end of the $K$-band.

Consistent with previous KPIC observations, orders 37--39, covering 1.94--2.09 $\mu$m, are heavily contaminated by telluric features and omitted from subsequent analysis. Initial cross-correlation analysis found that orders 35 and 36 do not meaningfully contribute to the planet detection due to the absence of strong spectral features. We therefore restrict our analysis to orders 31--34, spanning 2.22--2.49 $\mu$m, in order to speed up retrievals.

\subsection{Cross-correlation}

Propagating the reported period uncertainty from \citet{hebrard2016} to the observation epoch results in an uncertainty of approximately 12 hours in the conjunction time. We therefore began with a cross-correlation analysis in order to isolate the planet signal for subsequent retrieval. Rather than varying the planet RV semi-amplitude $K_p$ and the systemic velocity $v_{sys}$, we instead vary $K_p$ and the conjunction time, $\Delta \rm T_{conj}$. This better reflects existing knowledge of a system where the inclination is unknown and the orbital phase has significant uncertainties, but the systemic velocity is well-constrained by previous radial velocity observations. In particular, the short orbital period of \HDb\ creates a scenario where a change in orbital phase corresponding to just a few hours can dramatically change the velocity track of the planet, from nearly linear in the case of observations near conjunction to clearly sinusoidal for phases close to quadrature.

To create the $K_p - \Delta \rm T_{conj}$ plots, we projected out 6 principal components during detrending, and computed the cross-correlation rather than the log-likelihood. The \citet{brogi2019} log-likelihood function shows strong systematic effects in the $K_p - \Delta \rm T_{conj}$ space due to its explicit dependence on the model variance. As a result of the time series detrending and its replication in the forward modeling, the model variance is itself strongly dependent on the velocity shift of the planet, which changes with both $K_p$ and $\Delta \rm T_{conj}$. The cross-correlation coefficient is much less dependent on the variance of the planet models, making it a more appropriate choice for detection in this case. See \citet{finnerty2024} for a more extended discussion of these systematics in the $K_p-v_{sys}$ case.

\subsection{Atmospheric Retrieval}

We continue to use the retrieval framework described in \citet[Finnerty et al., submitted]{finnerty2023, finnerty2024}, with some minor updates which we summarize below. Our retrievals now use \petit\ version 3 \citep{prt:2019, prt:2020, Nasedkin2024}, which significantly improves the efficiency of the radiative transfer calculation. We also add a grey cloud opacity, following the \petit\ documentation: 
\begin{equation}
    \kappa_{\rm cld} = \kappa_{\rm 0,cld}\left(P/P_0\right)^{f_{\rm SED}}
\end{equation}
which is applied for $P > P_0$. We treat $P_0$, $\kappa_{\rm 0,cld}$, and $f_{\rm SED}$ as free parameters in the retrieval. 

Additionally, we have changed the PCA to be performed on the temporal axis of the data arrays, rather than the spectral axis. This provides a minor performance improvement, and does not change the resulting log-likelihood for small numbers of omitted components. The resulting principal component vectors are also somewhat easier to interpret, representing the way the observed spectrum evolves in time, rather than expressing the time series as a sum of component spectra with time-varying weights. 

Based on the ephemeris uncertainty discussed previously, we opt to update the nominal $\rm T_{conj}$ for the retrieval analysis to JD 2456531.2225 based on the mean offset in from the four models shown in Figure \ref{fig:kpdphi}. We then fit for a small offset in the conjunction time in the retrieval, rather than $v_{sys}$. This offset corresponds to a $\pm1.2$ hour shift in the conjunction time, which was chosen based on the range of offsets from Figure \ref{fig:kpdphi}.

Finally, we have changed the nested sampling implementation from \texttt{dynesty} \citep{speagle2020} to \texttt{MultiNEST} \citep{feroz2008, feroz2009, buchner2014, feroz2019}. Both packages produce similar final posteriors, but \texttt{MultiNest} is approximately $10\times$ faster, enabling retrievals in 12 hours of wall time with 8 CPU cores. We use 500 live points and a $\Delta \log z < 0.1$ convergence criteria. The final posteriors are consistent with runs using a larger number of live points and/or a stricter convergence criteria (e.g. 1000 or 1200 live points, $\Delta \log z < 0.01$), but the smaller number of live points and looser $\Delta \log z$ criteria significantly reduce the total run time. 

The equilibrium temperature of \HDb\ estimated with the system properties listed in Table \ref{tab:props} is $\sim$1940 K. This is close to the transition temperature where models predict dayside thermal inversions \citep[e.g.][]{baxter2020, roth2024}. We therefore choose a wide prior on the $\gamma$ parameter of the of the \citet{guillot2010} Pressure-Temperature ($P-T$) profile to permit either inverted or non-inverted atmospheres. While the internal temperature parameter $\rm T_{int}$ has not been well-constrained in previous retrievals of hot Jupiter atmospheres with KPIC data, we include it as a free parameter for completeness, and similarly include the planetary surface gravity $\log g$ as a free parameter. Tests using the \texttt{easyCHEM} \citep{molliere2017, lei2024} equilibrium chemistry calculator indicate the dominant metal species at the expected temperatures in the atmosphere of \HDb\ are CO, H$_2$O, Fe, and Mg, and that the vertical mixing profiles for these species are nearly constant with altitude for the non-inverted $P-T$ profiles favored in the initial cross-correlation analysis. We therefore treat all abundances as constant with altitude. 

For the molecular species, we use the opacity tables previously described in \citet{finnerty2023}. For H$_2$O, these tables used the \citep{polyansky2018} partition function and HITEMP 2010 linelist \citep{hitemp2010}. For both CO isotopologues we used the HITEMP 2019 lists \citep{hitemp2020} and the \citet{li2015} partition function. For Fe and Mg we use the \petit\ provided tables based on the Kurucz atomic line lists.

We ran retrievals omitting 4, 6, and 8 principal components during the data detrending process. We report the values from the 6 component retrieval in Section \ref{sec:res}. The marginalized medians and confidence intervals are consistent between all three retrievals, indicating the values we report are not strongly dependent on the number of PCA components. 

\section{Results}\label{sec:res}

\subsection{Cross-correlation Analysis}
For the initial cross-correlation analysis, we generated models based on the equilibrium temperature of \HDb\ and varying assumed compositions under chemical equilibrium. For the $P-T$ profile, we assumed $\log \kappa = -1$, $\log \gamma = -1$ for the non-inverted case and $\log \gamma = 1$ for the inverted case, $\log g = 3.3$, $\rm T_{int} = 100$ K, and $\rm T_{eq} = 1940$ K. The assumed cloud parameters were $\rm f_{SED} = 4$, $\log \kappa_{\rm 0,cld} = 1$, $\log \rm P_{cloud} = 0$. We assumed planetary $v\sin i = 3$ \kms, corresponding to tidally-locked rotation for a planetary radius of $1.2\rm R_J$, consistent with a moderately inflated hot Jupiter. For the molecular abundances, we used \texttt{easychem} \citep{molliere2017, lei2024} to compute equilibrium abundance profiles for H$_2$O, CO, Fe, and Mg assuming different atmospheric C/O ratios (0.3 or 0.8), metallicities (-0.3 or +0.3 dex relative to solar), and whether or not a thermal inversion is present, for a total of eight planet models. For the inverted models, we did not account for vertical variations in species abundances, although \texttt{easychem} predicts H$_2$O dissociates above the inversion and Fe and Mg are not present in significant quantities below the inversion. We instead took the respective lower/upper atmosphere abundances as constant. While the would create model mismatches in a retrieval analysis, it should not preclude  simple detection.

\subsubsection{Ephemeris and velocities}

\begin{figure*}
    \centering
    \includegraphics[width=1.0\linewidth]{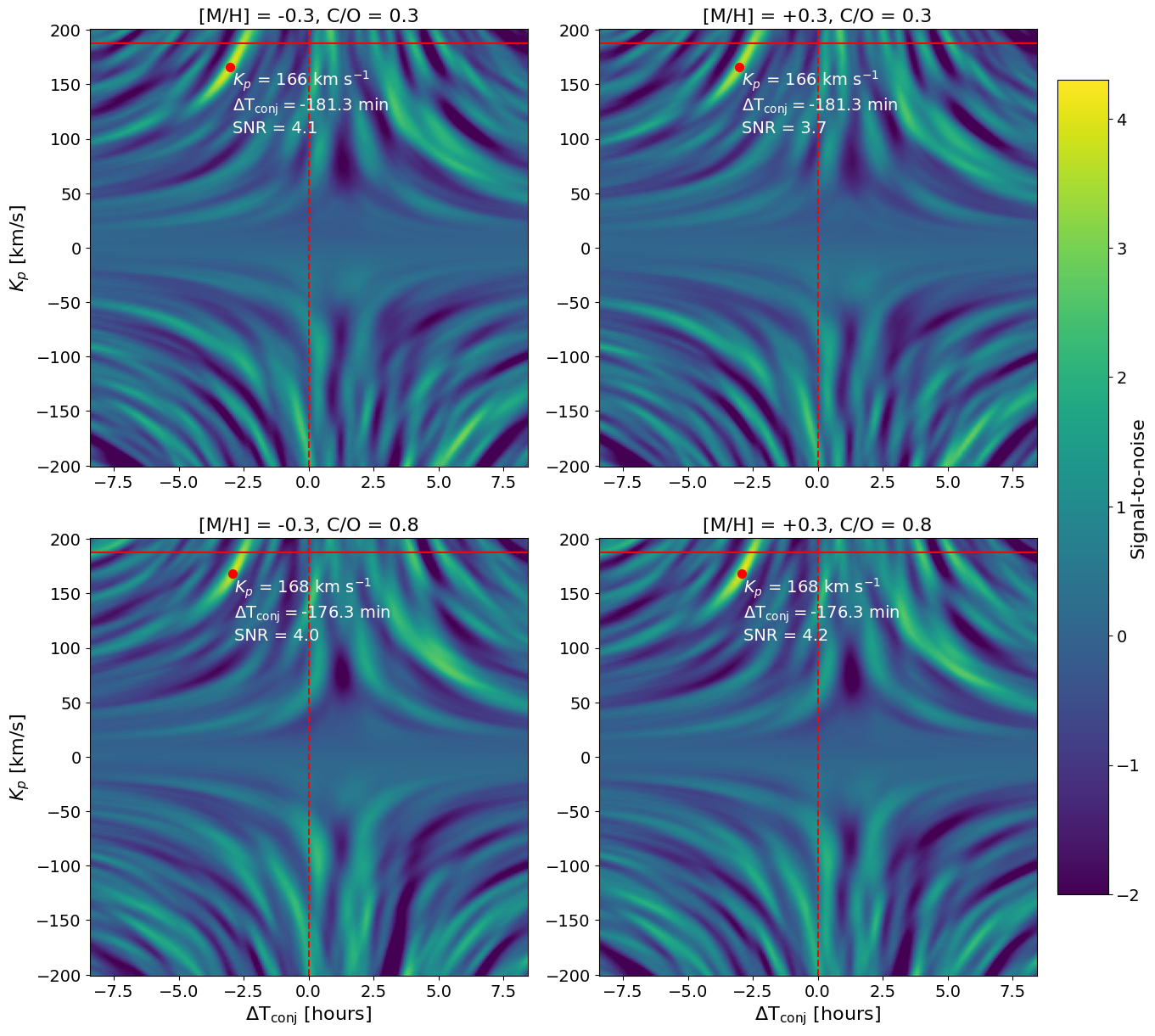}
    \caption{$K_p$ vs conjunction time offset the four non-inverted planet models with varying metallicity and C/O ratio. All four models show a strong feature consistent with a planet detection at similar $K_p$ and $\Delta \rm T_{conj}$. The peak values correspond to a conjunction time approximately 3 hours before the expectation from \citet{hebrard2016}, well within the propagated uncertainty in the conjunction time arising from uncertainty in the orbital period. The peak planet RV semi-amplitude is $\sim167$ \kms\ for all models, corresponding to an inclination of 64 degrees. The signal-to-noise was calculated by dividing the map by the standard deviation of the $K_p < -60$ region. All models are detected at similar strength.}
    \label{fig:kpdphi}
\end{figure*}

\begin{figure*}
    \centering
    \includegraphics[width=1.0\linewidth]{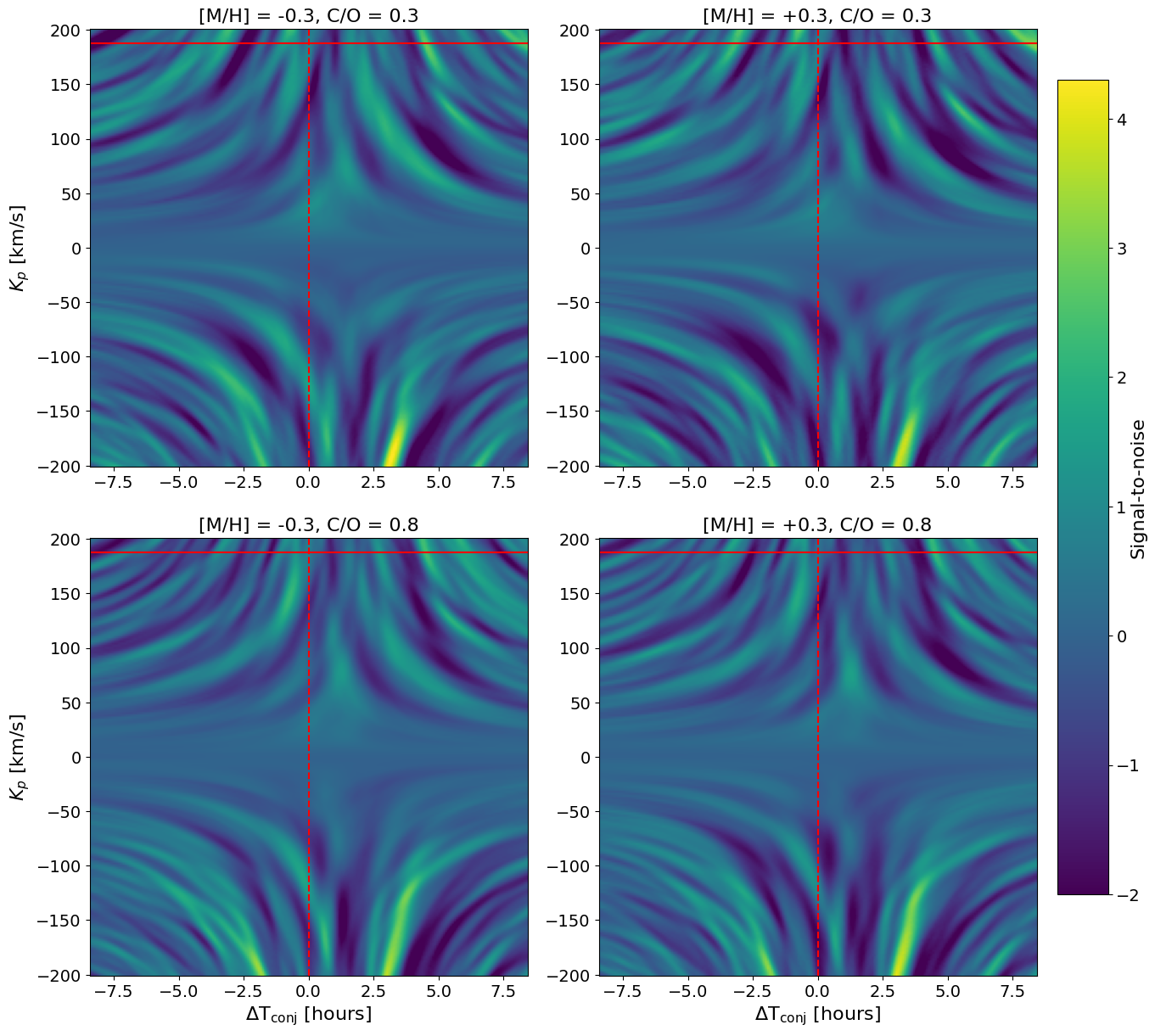}
    \caption{$K_p$ vs conjunction time offset the four inverted  planet models with varying metallicity and C/O ratio. None of these models give a potential planet detection in the region permitted by the prior ephemeris, indicating the detected planet flux is primarily arising from regions of the planetary atmosphere without a thermal inversion. The peak at $\rm T_{conj} = +2.5$ hours, $K_p =  -175$ \kms\ corresponds to a velocity track near transit/inferior conjunction, when the observed disk is dominated by the nightside of the planet. As this is incompatible with the previous ephemeris constraints, and we expect the nightside atmosphere to be non-inverted, we believe this feature arises from cross-talk with stellar residuals and is not related to the planetary atmosphere. }
    \label{fig:kpdphiinv}
\end{figure*}

\begin{figure*}
    \centering
    \includegraphics[width=0.4\linewidth]{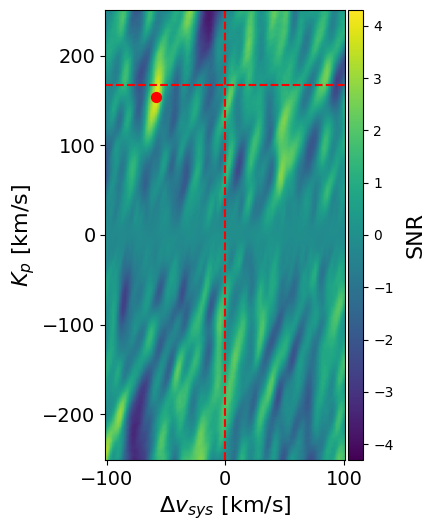}
    \includegraphics[width=0.4\linewidth]{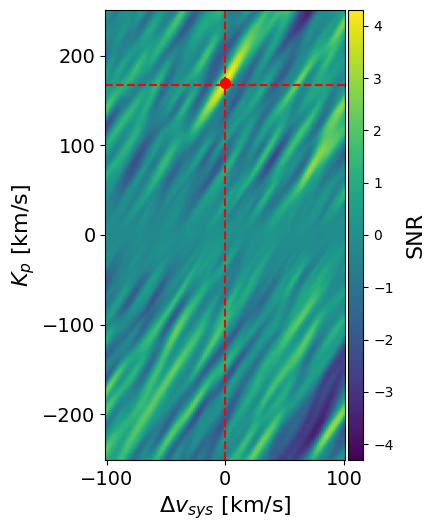}
    \caption{$K_p - v_{sys}$ diagrams for the \citet{hebrard2016} ephemeris (left) and the revised ephemeris based on Figure \ref{fig:kpdphi} (right). The Z = 0.3, C/O = 0.8, non-inverted model was used in both cases. The \citet{hebrard2016} ephemeris produces a detection of comparable strength, but with $\Delta v_{sys} = -60$ \kms, well outside the expected range of offsets for a planet with a well-constrained ephemeris.}
    \label{fig:kpvsys}
\end{figure*}

Figure \ref{fig:kpdphi} shows the computed $K_p - \Delta \rm T_{conj}$ plots for the four non-inverted planet models, while Figure \ref{fig:kpdphiinv} shows the four inverted models. Figure \ref{fig:kpvsys} shows the $K_p - v_{sys}$ diagram for a non-inverted model with both the \citet{hebrard2016} and our revised ephemerides for comparison.  All four non-inverted models show a $\rm SNR\sim4$ feature at approximately $\Delta \rm T_{conj} = -180$ minutes, $K_p = 167$ \kms. None of the inverted models produce features of comparable strength, and all of the inverted models are consistent with a non-detection, indicating that the observed planet flux is primarily arising from regions without a thermal inversion. Based on the reported semi-major axis and orbital period from \citet{hebrard2016}, the maximum value for $K_p$, corresponding to a 90 degree inclination, is 188$\pm$4 \kms, so the $K_p = 167$ \kms\ value from the non-inverted models is consistent with a detection of the non-transiting \HDb. The $\Delta \rm T_{conj}$ offset is consistent with the expected uncertainty from propagating the \citet{hebrard2016} uncertainty in the orbital period to the observation epoch. While there are slight shifts in the location of the cross-correlation peak for different non-inverted models, these shifts are relatively small and consistent with the strong degeneracy between $K_p$ and $\Delta \rm T_{conj}$. 

We use the scale of the $K_p$ and $\rm \Delta T_{conj}$ offsets to inform the priors used for atmospheric retrieval. We update the nominal conjunction time to $\rm T_{conj} = \rm\ JD\ 2456531.2225$ and the nominal $K_p$ to 167 \kms. We use these values for the remaining analysis. 

\subsubsection{Molecular detections}

\begin{figure*}
    \centering
    \includegraphics[width=0.9\linewidth]{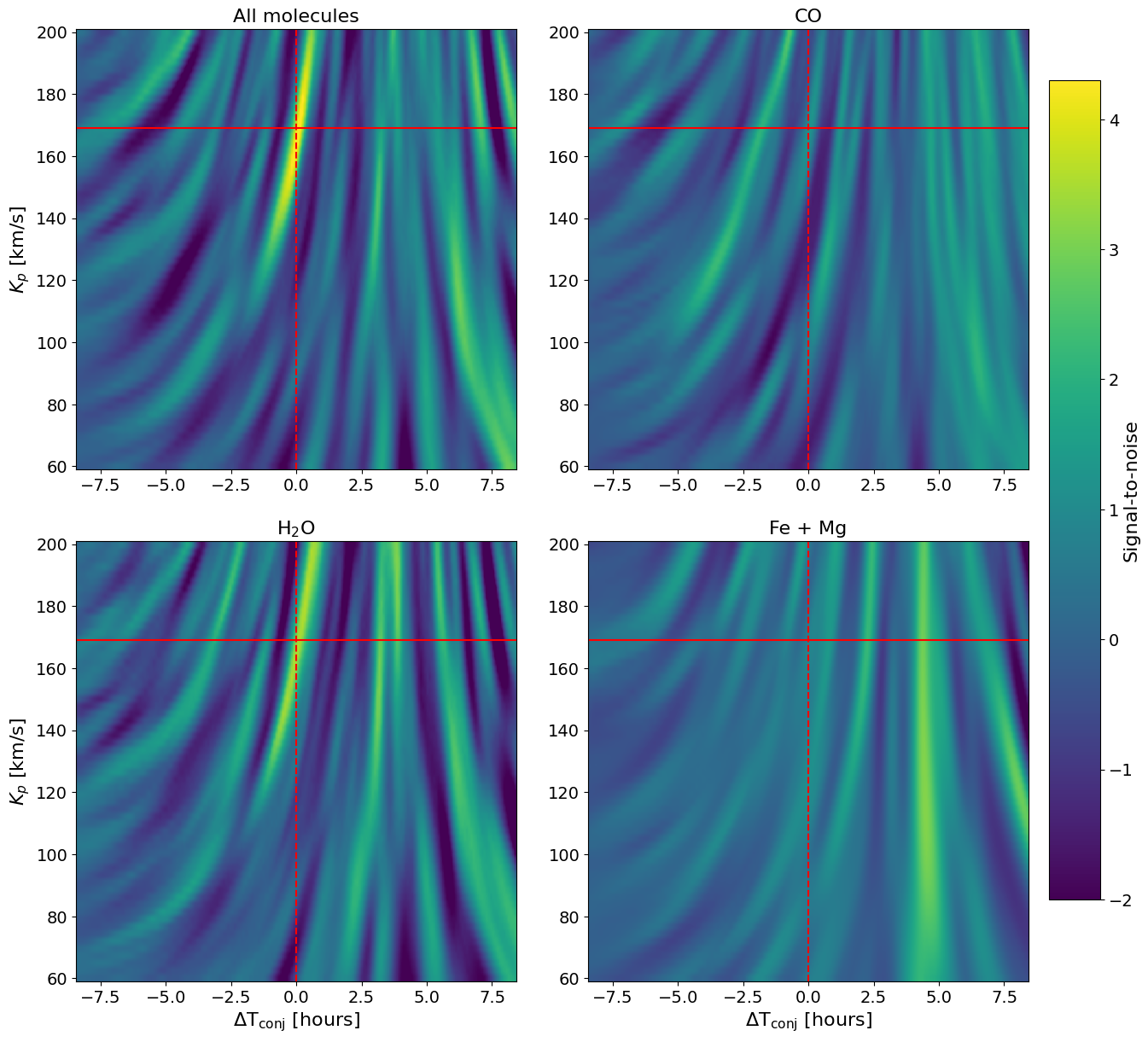}
    \caption{$K_p$ vs conjunction time offset for models of all species, CO, H$_2$O, and Fe+Mg, with abundances based on the non-inverted [M/H] = 0.3, C/O = 0.8 model which produced the strongest detection in Figure \ref{fig:kpdphi}. The conjunction time has been updated based on the measured offsets from Figure \ref{fig:kpdphi}, so planetary features should appear at approximately $K_p = 167$ \kms, $\Delta \rm T_{conj}$ = 0 hours. Only H$_2$O is confidently detected. A weak feature is present at the same location with the CO model, but does not dominate the $K_p - \Delta T_{\rm conj}$ space and does not constitute an independent detection. The refractory species are not detected.}
    \label{fig:mols}
\end{figure*}

Plots of $K_p - \Delta \rm T_{conj}$ for the the [M/H] = 0.3, C/O = 0.8 model, and individually for CO, H$_2$O, and Mg+Fe, are shown in Figure \ref{fig:mols}. We use the same $P-T$ parameters throughout, and base abundances on the [M/H] = 0.3, C/O = 0.8 model, which produced the strongest detection in Figure \ref{fig:kpdphi}. 

Only H$_2$O produces a significant detection in isolation. The CO model shows a weak peak consistent with the planet velocity from the H$_2$O model, but other features dominate the cross-correlation space. The refractory species do not produce any discernible peak near the expected planet velocity, and the cross-correlation space for these species is dominated by a feature in the stellar rest frame as a result of Fe and Mg lines in the stellar atmosphere. However, the all-molecule case is better detected than any individual molecules. This suggests that CO and the refractory species may be contributing some improvement to the overall detection, even if they are not independently detected in the absence of H$_2$O.

The absence of a significant CO detection appears to be at odds with the slight preference for high C/O from the initial cross-correlation analysis, shown in Figure \ref{fig:kpdphi}. However, we note that the increase in detection strength over the low C/O models in Figure \ref{fig:kpdphi} is marginal, and estimating the noise level for the $K_p - \Delta \rm T_{conj}$ diagram signal-to-noise by taking the off-peak cross-correlation variance has some associated uncertainty. Furthermore, the large number of H$_2$O lines in the $K$-band leads to a pseudocontinuum in the planet model which differs significantly from a blackbody, which may lead to worse detections when using single-molecule templates omitting H$_2$O \citep{finnerty2024}. As a result, we rely on the Bayesian retrieval analysis to constrain the chemical composition, rather than the cross-correlation.  

\subsection{Retrieval Analysis}
Priors, maximum-likelihood parameters, and retrieved medians with $\pm34$\% confidence intervals are presented in Table \ref{tab:priors}. Our free-chemistry approach does not directly fit for [C/H], [O/H], or C/O, but we use the retrieved posteriors to construct posteriors for these parameters, and report the corresponding values in Table \ref{tab:priors} as derived parameters. We present the results of the retrieval omitting 6 principal components. The medians and confidence intervals for the 4 and 8 component cases did not significantly differ from the values reported in Table \ref{tab:priors}. 

\begin{deluxetable*}{ccccc}
    \tablehead{\colhead{Name} & Symbol  & \colhead{Prior} & \colhead{Retrieved Max-L} & \colhead{Retrieved Median}}
    \startdata
        log infrared opacity [$\rm cm^{2} g^{-1}$] & $\log \kappa$ &  Uniform(-3, 3) &  -1.9 & $-1.2^{+1.0}_{-0.9}$ \\
        log infrared/optical opacity & $\log \gamma$  &  Uniform(-3, 3) & -2.5 & $-1.7^{+0.7}_{-0.7}$ \\
        Intrinsic temperature [K] & $\rm T_{int}$ & Uniform(50, 1000) & 410 & $<800$ \\
        Equilibrium temperature [K] & $\rm T_{equ}$ & Uniform(800, 3000) &  1100 & $1400^{+400}_{-300}$  \\
        log surface gravity [cgs] & $\rm \log g$ & Uniform(2,4) & 2.2 & $3.0^{+0.6}_{-0.6}*$ \\
        Cloud opacity index & $\rm f_{SED}$ & Uniform(0,8) & 7.9 & $4.8^{+2.0}_{-2.1}*$ \\
        Cloud base opacity & $\log \kappa_0$ & Uniform(0,3) & 1.5 & $1.5^{+1.0}_{-0.9}*$ \\
        Cloud base pressure & $\log \rm P_{cloud}$ & Uniform(-4,2) & -0.3 & $>-1.7$  \\
        $K_{\rm p}$ offset [\kms] & $\Delta K_{\rm p}$  & Uniform(-40, 40) & 12.8 & $17.9^{+10.6}_{-13.2}$ \\
        Conjunction time offset [hours] & $\Delta \rm T_{conj}$ & Uniform(-1.2, 1.2) &  0.27 & $0.47^{+0.24}_{-0.28}$ \\
        Rotational velocity [\kms] & $v_{\rm rot}$ & Uniform(0, 15) & 5.8 & $7.4^{+3.1}_{-3.5}*$   \\
        log H$_2$O mass-mixing ratio & log H$_2$O  &  Uniform(-12, -0.5) & -3.9 & $-3.9^{+0.8}_{-0.5}$  \\
        log CO mass-mixing ratio & log CO &  Uniform(-12, -0.5) &  -4.7 & $<-3.7$ \\
        log Fe mass-mixing ratio & log Fe & Uniform(-12,-1) &  -4.1 & $-6.5^{+3.3}_{-3.4}*$ \\
        log Mg mass-mixing ratio & log Mg & Uniform(-12,-1) &  -1.6 & $-6.6^{+3.4}_{-3.2}*$ \\
        log $\rm ^{13}CO/^{12}CO$ & $\rm \log ^{13}CO_{rat}$ &  Uniform(-8, -0.5) & -5.1 & $-4.2^{+2.3}_{-2.2}*$  \\ 
        log H mass-mixing ratio & $\log \rm all H$ &  Uniform(-0.4, -0.05) & -0.2 & $-0.2^{+0.1}_{-0.1}*$  \\ 
        Scale factor & scale & LogNormal(0, 0.2) & 0.4 & $0.2^{+0.2}_{-0.2}$
        \smallskip \\
         & & \textbf{Derived Parameters} & \\
        \hline
        Carbon/oxygen ratio & C/O & - & 0.1 &  $<0.2$  \\
        % Refractory/volatile ratio & [R/V] & - & & \\
        Carbon abundance & [C/H] & - & -2.1 & $<-1.1$ \\
        Oxygen abundance & [O/H] & - & -1.3 & $-1.3^{+0.9}_{-0.5}$ \\
        Volatile abundance & [(C+O)/H] &  - & -1.5 & $-1.5^{+0.9}_{-0.5}$ \\
    \enddata 
    \caption{List of parameters, priors, and results for atmospheric retrievals. The error bars on the retrieved medians correspond to the 68$\% / 1\sigma$ confidence interval.  Limits are given at 95\% confidence, omitting the secondary mode for the C/O ratio}. A $*$ indicates the marginalized posterior for a parameter is unconstrained (i.e. the marginalized posterior spans the full range of the prior). In addition to these priors, we required that the atmospheric temperature stay below 6000 K at all pressure levels. The full corner plot is included in Appendix \ref{app:corner}.
    \label{tab:priors}
\end{deluxetable*}

\subsubsection{Kinematics}

The retrieved $K_p$ and $\Delta \rm T_{conj}$ are slightly offset compared with the values from the cross-correlation analysis shown in Figure \ref{fig:kpdphi} and \ref{fig:mols}, but are consistent with the degeneracy between these parameters. The prior on $\Delta \rm T_{conj}$ corresponds to a $\pm1.2$ hour change in the conjunction time, based on the apparent extent of the planet peak in \ref{fig:kpdphi}, and $\Delta \rm T_{conj}$ is well-constrained to within $\sim25$ minutes. %This is consistent with the scale of the change in the planet velocity resulting from changes in $\Delta \rm M$, which is approximately $\pm5$ \kms, compared to a velocity resolution of 9 \kms\ for NIRSPEC \citep{nirspecupgrade2}. 

The retrieval provides a weak constraint on the planetary rotational velocity,  $v\sin i = 7.4^{+3.1}_{-3.5}$ \kms. This is comparable to the spectrograph resolution, corresponding to a $\sim$30\% broadening of the planet lines compared with the instrumental profile alone. The expected $v\sin i$ for tidally-locked rotation is $\sim$3 \kms, leaving $\sim$5 \kms\ of unexplained broadening. This could be consistent with apparent broadening from a broadly-distributed westward-flowing wind pattern, similar to that reported for other hot and ultra-hot Jupiters \citep[e.g.][]{pai2022, gandhi2023}. However, the poor quality of the $v\sin i$ constraint and uncertainties in the true LSF preclude a definitive physical interpretation of this value. %, and are commonly seen in Global Circulation Model (GCM) results. 

We used the same line-spread treatment as that developed in Finnerty et al. (subm.), and have the same systematic uncertainty in the overall LSF  width at the $\sim$10\% level. While the retrieved broadening is somewhat larger than we believe could be explained solely by a systematic error in the LSF, we are cautious about making specific quantitative or physical interpretations of the retrieved broadening, particularly given the broad marginalized posterior for $v\sin i$. Higher-resolution observations, particularly from instruments with robust LSF calibration (as is planned for Keck/HISPEC, \cite{hispec}), are necessary to quantitatively interpret line broadening in planetary atmospheres. A $v\sin i$ of 7 \kms\ would be $>2\times$ the spectral resolution of Keck/HISPEC, and should be easily distinguishable.  

\subsubsection{Pressure-temperature profile and clouds}\label{ssec:thermal}

\begin{figure*}
    \centering
    \includegraphics[width=0.95\linewidth]{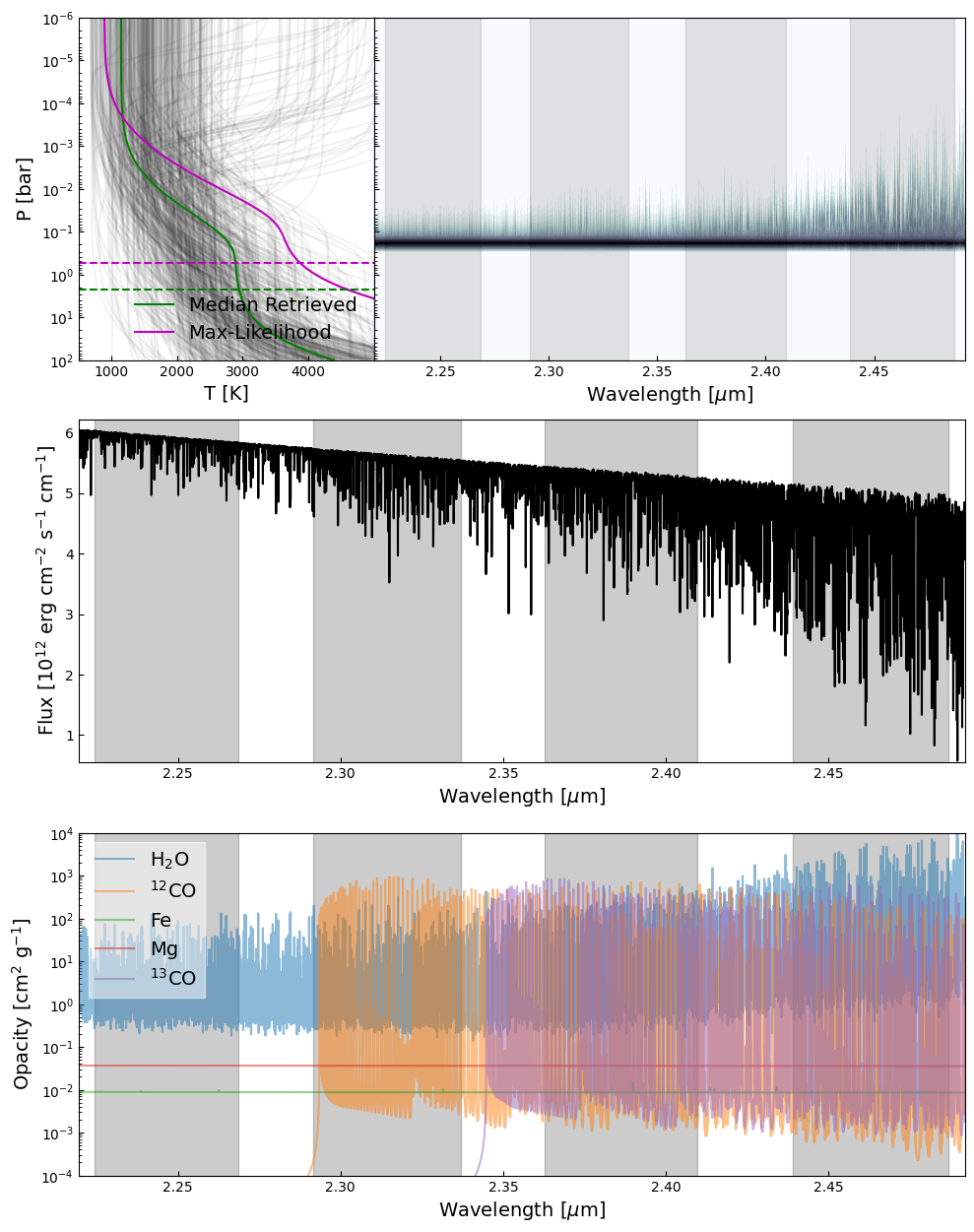}
    \caption{Retrieved $P-T$ profile (top left), maximum-likelihood emission contribution function (top right), maximum-likelihood spectrum (middle), and species opacities (bottom). In the $P-T$ plot, the median retrieved profile is shown in green, and the maximum likelihood profile in purple. The cloud pressures are indicated by horizontal dashed lines. The black lines are the $P-T$ profiles from 500 draws from the retrieved posteriors. In the emission contribution function, spectrum, and opacity panels, the observed orders are shaded in grey. Most of the emission arises between 1 bar and 30 mbar, where the $P-T$ profile begins to cool, and is significantly above the cloud pressure, indicating the atmosphere is clear at the pressures probed by these $K$-band observations. The maximum-likelihood spectrum shows weak CO features in addition to H$_2$O, and the non-detection of refractory species is consistent with the extremely low opacity of these species in comparison to CO and H$_2$O}
    \label{fig:PT}
\end{figure*}

The retrieved $P-T$ profile and emission contribution function are shown in Figure \ref{fig:PT}. The retrieved $P-T$ profile is non-inverted and somewhat colder than expected based on the equilibrium temperature of 1940 K. This is consistent with previous KPIC retrievals \citep[Finnerty et al., submitted]{finnerty2023, finnerty2024}. As with these previous results, the overall scaling factor applied to the planet model is $>$1, compensating for the lower temperature. Broader wavelength coverage that covers a significant change in the planet continuum flux (such as the region surrounding the blackbody peak) would break this degeneracy. 

 % This can be clearly seen in the difference between the median and maximum-likelihood retrieved $P-T$ profiles in Figure \ref{fig:PT}. The median profile is 200 K hotter than the maximum-likelihood profile, but this is compensated by a doubling of the scale factor applied to the final spectrum. 

Of the cloud parameters, only the pressure is constrained. The retrieved pressure is quite deep, with the posterior peaking around 2.5 bar, well below the bottom of the emission contribution function shown in Figure \ref{fig:PT}. At these pressures, the final impact on the outgoing planet flux is negligible, so $\rm f_{SED}$ and $\kappa_{0,cld}$ are unconstrained, and the atmosphere over the pressure range probed by the $K$-band observations is effectively clear.  

\subsubsection{Chemical composition}\label{ssec:chem}

Consistent with the cross-correlation analysis, the retrieval provides a bounded constraint only for the H$_2$O abundance, $\log \rm H_2O = -3.9^{+0.8}_{-0.5}$. While the maximum likelihood value for the CO mass fraction is $\log \rm CO = -4.7$, the retrieval does not provide a lower bound on the CO abundance. High CO abundances are strongly excluded ($\log CO_{MMR} < -3.7$ at 95\% confidence). As a result, while the maximum-likelihood C/O ratio is 0.1, the median value from the retrieval is consistent with 0. This outcome could be explained by a CO abundance at or just below the detection limit of our observations. The CO isotopologue ratio is entirely unconstrained, consistent with the non-detection of the primary isotopologue.

The refractory species abundances are entirely unconstrained. While other $K$-band observations \citep[e.g.][]{ramkumar2023} have been able to constrain Fe abundances in hot Jupiter atmospheres, we suspect that the lower spectral resolution of NIRSPEC is insufficient to make a detection based on the small number of accessible Fe and Mg lines (see Figure \ref{fig:PT}, bottom panel). 

\section{Discussion}\label{sec:disc}

\subsection{Ephemeris revision and true mass}

Our analysis indicates the inferior conjunction of \HDb\ prior to our observation occurred approximately 3 hours before the expected time based on the \citet{hebrard2016} ephemeris. This is substantially larger than their reported uncertainty in $\rm T_{conj}$, but is well within the uncertainty resulting from propagating the orbital period uncertainty to the observed epoch. This suggests that uncertainty in the orbital period of \HDb\ is the primary contributor to the observed change in the conjunction time. A reduction of 0.1 minutes in the orbital period would entirely explain the change in conjunction time, and is compatible with the 0.4 minute period uncertainty reported in \citet{hebrard2016}. While our limited phase coverage prevents accurate revision of the orbital period, the revised conjunction time from the retrieval, $\rm T_{conj} = 2456531.242^{+0.010}_{-0.012}$, will enable more precise timing of follow-up observations for the next several years. This demonstrates that high-resolution cross-correlation techniques do not necessarily require precise ephemeris knowledge, as long as a sufficiently large velocity shift is observed for the planet signal to be clearly detected in the $K_p - \Delta \rm T_{conj}$ space.

\HDb\ is non-transiting, so the orbital inclination and true mass were not known \textit{a priori}. As the measured value of $K_p$ is the radial projection of the total orbital velocity, it can be used to calculate the inclination as fellows:
\begin{equation}
    K_p = K_{p,max} \sin i = \frac{2\pi a}{T}\times \sin i
\end{equation}

The semi-major axis $a$ is not directly observable, but can be replaced using Kepler's third law to give:

\begin{equation}
    K_p = \sin i \times \left(\frac{2\pi G M_*}{T}\right)^{1/3}
\end{equation}

Using the stellar mass and orbital period from \citet{hebrard2016}, we obtain $K_{p,max} = 187.9^{+4.4}_{-4.7}$ \kms, with the uncertainty dominated by the uncertainty in the stellar mass. We can then solve for the inclination:

\begin{equation}
    \sin i = K_p/K_{p,max} = 0.98\pm0.07
\end{equation}

Which gives $i = 78^{\circ+12}_{-12}$, restricting to $\sin i < 1$. While $\sin i$ is strictly $<1$, we have neglected the degeneracy between $K_p$ and $K_{p,max}$ when propagating uncertainties, so $\sin i > 1$ appears to be compatible with the uncertainties despite being non-physical. Larger stellar masses increase $K_{p,max}$, permitting higher values of $K_p$, and inclusion of $\sin i > 1$ in the statistical error therefore suggests the stellar mass may be somewhat underestimated. A more informative upper bound can be obtained from the lack of a reported transit, which requires $i < 80^\circ$ for the reported stellar parameters from \citet{hebrard2016}. We therefore adopt $i = 78^{\circ +2}_{-12}$ as the inclination of the system. We can then determine the true planet mass by dividing the minimum mass obtained from radial velocity measurements by $\sin i$, yielding $\rm M_{true} = 1.23\pm0.10 M_J$, including the $M\sin i $ uncertainty from \citet{hebrard2016}. %The uncertainty in this value is dominated by the reported uncertainty in $M\sin i$ from \citet{hebrard2016}. 

This result makes \HDb\ one of a handful of non-transiting planets with a known true mass determined via high-resolution cross-correlation techniques \citep{brogi2012, brogi2014, birkby2017, guilluy2019}. This is currently the only technique which can retrieve atmospheric properties for close-in, non-transiting planets. 

\subsection{Thermal structure}

The retrieved posterior for the $\gamma$ parameter, which determines whether the $P-T$ profile has a thermal inversion, strongly prefers a non-inverted structure, with a weak tail permitting inversions. This is potentially consistent with a weak, localized thermal inversion close to the substellar point marginally contributing to the observed flux. The apparently low C/O ratio of \HDb\ significantly complicates efforts to confirm any inversion. \texttt{easyCHEM} chemical equilibrium models predict H$_2$O should rapidly dissociate in the thermal inversion, leaving CO as the only species with significant $K$-band opacity in the inverted region of the atmosphere. If the CO abundance is very low, as the retrieval analysis prefers, this may lead to a featureless $K$-band spectrum from any inverted region, preventing detection of that region with HRCCS.

The relatively edge-on retrieved inclination of \HDb\ suggests the lack of an inversion is not simply an issue of viewing geometry. For a low-inclination (e.g. $i<60^\circ$), our post-conjunction observations would be much more sensitive to morning longitudes and high latitudes compared with evening/equatorial regions. This could significantly decrease the visibility of a localized inversion, which models predict would occur near equatorial latitudes and substellar or afternoon longitudes \citep{roth2024}. However, our analysis prefers $i \sim 78^\circ$ for \HDb, and should therefore still be sensitive to equatorial latitudes, making it difficult to miss a thermal inversion of significant extent as a result of viewing geometry alone.

Observationally, thermal inversions are routinely seen in planets with equilibrium temperatures $>2200$ K \citep[e.g.][Finnerty et al., subm.]{finnerty2023, ramkumar2023}, but not seen for temperatures $<1800$ K \citep[e.g.][]{line2021, pelletier2021, finnerty2024}. Beyond these observations of \HDb, relatively few objects have been characterized in the $1800-2200$ K range where this transition occurs. This makes \HDb\ a potentially important object for understanding when and how thermal inversions form in exoplanet atmospheres.

For equilibrium temperatures $>2500$ K, circulation models predict the presence of thermal inversion regardless of chemical composition \citep{lothringer2018}. At lower temperatures, inversions are strongly associated with the presence of metal oxide species, particularly TiO and VO \citep[e.g.][]{fortney2008}. The abundances of these species can be significantly impacted by cold trapping \citep{hubeny2003}, which may lead to localized inversions existing only where these species are in the gas phase. Nightside cold trapping of Ti has been reported in the UHJ WASP-76~b \citep[$\rm T_{eq} = 2200$ K][]{pelletier2023}, but the details of this process are difficult to model, and the temperature at which thermal inversions form in typical hot Jupiters is unclear. Modeling by \citet{roth2024} using 3D GCMs suggests the transition temperature to thermally-inverted $P-T$ profiles is $\sim1800$ K in the presence of TiO/VO, while previous 1D modeling by \citet{baxter2020} preferred an onset around 1700 K, but models omitting TiO/VO do not show inversions until substantially higher temperatures. These models predict that \HDb, with an equilibrium temperature of $\sim1900$ K, should have a dayside inversion only if significant quantities of TiO/VO are present.

Given the low retrieved metallicity of \HDb, there may simply be insufficient TiO/VO in the atmosphere to produce a thermal inversion, especially if Ti is being cold-trapped on the nightside. TiO has substantially greater opacity in the $H$ and $J$ bands compared with $K$, which should allow the upcoming Keck/HISPEC instrument \citep{hispec} to either detect TiO or place a meaningful upper limit on its abundance. The revised ephemeris from this work will make it easier to obtain broad phase coverage in followup observations for phase-resolved retrievals. Such analysis will allow the partial inversion scenario to be definitively ruled out and will offer greater clarity into potential cold trapping impacting the TiO abundance. 

\subsection{C/O and metallicity}

\begin{figure}
    \centering
    \includegraphics[width=1.0\linewidth]{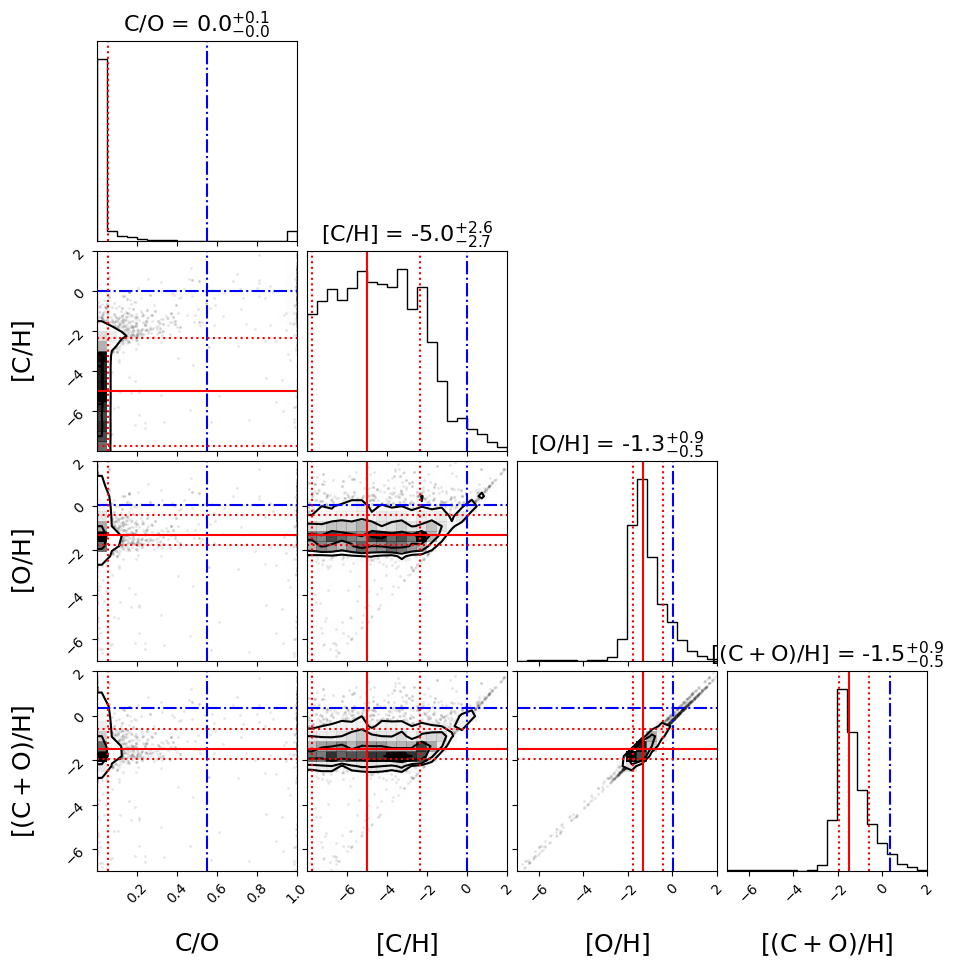}
    \caption{Corner plot showing the retrieved C/O, [C/H],[O/H], and [(C+O)/H] posteriors, computed from the retrieved abundances for H$_2$O and CO. Stellar values are indicated in dash-dot blue, while the retrieved median and $\pm34$\% confidence intervals are indicated in solid and dotted red, respectively. While the oxygen abundance is sub-solar/stellar at $\sim1\sigma$, the carbon abundance is only an upper limit, leading to an upper limit on the C/O ratio. These results are consistent with a metal- and carbon-poor atmosphere for \HDb.}
    \label{fig:cochplot}
\end{figure}

Derived posteriors for C/O, [C/H], [O/H], and [(C+O)/H] are shown in Figure \ref{fig:cochplot}. The oxygen abundance is sub-solar by $\sim$1.3 dex at just over $1\sigma$ significance. The retrieval provides only an upper limit on the atmospheric mass fraction for CO, $\log CO < -3.7$ at 95\% confidence. There is a weak, kinematically offset mode with very high CO and low H$_2$O abundance, which can be seen at the extreme right of the C/O posterior in Figure \ref{fig:cochplot}. As this mode is offset in both $K_p$ and $\rm T_{conj}$, we omit these points when estimating the upper limit on the C/O ratio. This leads to an upper limit on the C/O ratio, $\rm C/O < 0.2$ at 95\% confidence, while the maximum-likelihood value is C/O = 0.1 and the retrieved median is consistent with a nearly carbon-free atmosphere. The Bayes factor comparing the maximum-likelihood model with the same model omitting CO is only 2.9, consistent with a marginal preference for CO. This suggests the atmospheric CO abundance may be slightly below our detection threshold. Future observations with Keck/HISPEC \citep{hispec} will offer significant improvements in both instrument throughput and spectral resolution, which may enable precise constraints on the CO abundance and C/O ratio even in low-metallicity atmospheres. The continuous wavelength coverage of HISPEC may be particularly valuable in this case, as HISPEC observations will capture the full R and P branches associated with the 2.3$\mu$m CO bandhead, greatly increasing the number of observed CO lines compared with KPIC.

The low-C/O ratio, low-metallicity atmospheric composition our retrievals prefer for \HDb\ has not previously been observed in an exoplanet atmosphere, and contradicts model predictions that these parameters should be inversely correlated \citep{espinoza2017, cridland2019, khorshid2022}. In these models, oxygen enrichment usually arises from late planetesimal accretion, which should also increase the atmospheric metallicity. Low metallicity and low C/O ratio would require the accretion of significant quantities of oxygen-rich gas, but protoplanetary disk gas is generally carbon-rich as a result of condensation \citep{oberg2011}. \citet{chachan2023} finds that the region just beyond the H$_2$O snow line has substantially enriched H$_2$O ice relative to the rest of the disk, which could lead to an oxygen-rich atmosphere, although we would still expect a stellar or super-stellar metallicity in this scenario.

HRCCS observations have difficulty measuring absolute abundances due to the loss of continuum information during the data processing. As a result, we cannot conclusively rule out a stellar abundance of oxygen, and the C/H posterior similarly has a tail that extends to values compatible with the solar carbon abundance (see Figure \ref{fig:cochplot}). This alone could resolve the anomalous apparent composition of \HDb.  Furthermore, our estimate of the atmospheric metallicity is based only on abundances of volatile species. Inferring bulk metallicity from these species alone can be misleading due to either atmospheric effects, including oxygen bonding to refractory species \citep{lodder2003}, and cold trapping in the planetary atmosphere \citep{spiegel2009, pelletier2023}, or due to formation effects, including pebble drift and volatile trapping, which can alter the intrinsic refractory/volatile ratio \citep{booth2017, ligterink, lothringer2021, chachan2023}. Obtaining refractory species abundances can potentially resolve several of these scenarios by constraining the solid accretion history of a planet \citep{lothringer2021, chachan2023}, but will require additional observations in the case of \HDb. 

This unusual composition for \HDb\ is consistent with the emerging picture of significant chemical diversity in the hot Jupiter population \citep{kempton2024, wiser2024}. Reported metallicities for various exoplanets range from very high \citep[e.g.][]{bean2023} to moderately super-solar \citep[e.g.][]{finnerty2024, fu2024} to sub-solar \citep[e.g.][]{line2021, august2023, weinermansfield2024}. Similarly, reported C/O ratios range from substantially sub-stellar \citep[e.g.][]{finnerty2024, fu2024} to significantly carbon-enriched (Finnerty et al. submitted), although in the case of ultra-hot Jupiters constraining the C/O ratio is complicated by the impact of molecular dissociation \citep{brogi2023, ramkumar2023, gandhi2024}. This diversity suggests a variety of formation and evolutionary processes shape the hot Jupiter population, motivating further observations and improvements in retrieval techniques to incorporate a broader range of data, such as joint optical/infrared and high/low resolution retrievals. 

\section{Summary and Conclusions}\label{sec:conc}

We detected atmospheric H$_2$O absorption from the non-transiting hot Jupiter \HDb. Cross-correlation analysis indicates the conjunction time is 2.5 hours earlier than the published ephemeris, likely due to the accumulation of uncertainty in the orbital period over nearly a decade. The planet radial velocity semiamplitude is $K_p = 185^{+11}_{-13}$\kms, giving an orbital inclination of $78^{\circ+2}_{-12}$ and a true planet mass of 1.23$\pm0.10\rm\ M_J$, with an upper inclination limit imposed by the absence of a transit detection. This is the first non-transiting hot Jupiter detected with KPIC and the first atmospheric detection of \HDb. 

While the equilibrium temperature of \HDb\ is in the transition regime between non-inverted and inverted atmospheres, our analysis strongly prefers a non-inverted $P-T$ profile. This is unlikely to be solely the result of viewing geometry given the high inclination of \HDb, suggesting that the lack of an inversion may be a result of low bulk metallicity leading to low abundances of inversion-causing species such as TiO and VO. Additional observations with broader wavelength coverage would greatly improve sensitivity to these species and directly test this explanation. \HDb\ is potentially a key benchmark object for understanding the transition between inverted and non-inverted exoplanet atmospheres.

Retrieval analysis provides a bounded constraint on the H$_2$O mass-mixing ratio $\log \rm H_2O = -3.9^{+0.8}_{-0.5}$ and a 95\% upper limit on the CO mass-mixing ration $\log \rm CO < -3.7$. This places an upper limit on the C/O ratio $\rm C/O < 0.2$ at 95\% confidence, and suggests a sub-solar atmospheric metallicity. This combination of low metallicity and low C/O ratio is unusual in exoplanet atmospheres and has not been seen previously in high-resolution retrievals, but is consistent with other findings suggesting a chemically diverse hot Jupiter population. Future observations with Keck/HISPEC will have wider wavelength coverage and improved spectral resolution, which may be sufficient to detect CO in \HDb and establish a bounded constraint on the C/O ratio. Such observations may also provide constraints on refractory species such as iron or magnesium, which may clarify the formation history of \HDb\ through constraining the refractory-to-volatile ratio. 

Finally, observations with higher spectral resolution and/or wider phase coverage may enable detection of phase-dependent velocity shifts relative to a Keplerian orbit. Such shifts would offer insight into atmospheric circulation patterns, which are particularly interesting for non-transiting planets such as \HDb, as the observed emission probes higher latitudes compared with transiting planets. In the future, high-resolution spectroscopy of non-transiting planets may be a unique tool for probing atmospheric circulation patterns in polar regions of hot giant planets.

\begin{acknowledgments}
We thank the anonymous referee whose thoughtful comments improved the quality of this paper. 
L. F. is a member of UAW local 4811. L.F. acknowledges the support of the W.M. Keck Foundation, which also supports development of the KPIC facility data reduction pipeline. The contributed Hoffman2 computing node used for this work was supported by the Heising-Simons Foundation grant \#2020-1821. Funding for KPIC has been provided by the California Institute of Technology, the Jet Propulsion Laboratory, the Heising-Simons Foundation (grants \#2015-129, \#2017-318, \#2019-1312, \#2023-4597, \#2023-4598), the Simons Foundation (through the Caltech Center for Comparative Planetary Evolution), and the NSF under grant AST-1611623. D.E. acknowledges support from the NASA Future Investigators in NASA Earth and Space Science and Technology (FINESST) fellowship under award \#80NSSC19K1423, as well as support from the Keck Visiting Scholars Program (KVSP) to install the Phase II upgrades for KPIC. J.X. acknowledges support from the NASA Future Investigators in NASA Earth and Space Science and Technology (FINESST) award \#80NSSC23K1434.

This work used computational and storage services associated with the Hoffman2 Shared Cluster provided by UCLA Institute for Digital Research and Education’s Research Technology Group. L.F. thanks Briley Lewis for her helpful guide to using Hoffman2, and Paul Molli\`ere for his assistance in adding additional opacities to petitRADTRANS. 

The data presented herein were obtained at the W. M. Keck Observatory, which is operated as a scientific partnership among the California Institute of Technology, the University of California and the National Aeronautics and Space Administration. The Observatory was made possible by the generous financial support of the W. M. Keck Foundation. W. M. Keck Observatory access was supported by Northwestern University and the Center for Interdisciplinary Exploration and Research in Astrophysics (CIERA). The authors wish to recognize and acknowledge the very significant cultural role and reverence that the summit of Mauna Kea has always had within the indigenous Hawaiian community.  We are most fortunate to have the opportunity to conduct observations from this mountain. 

This research has made use of the NASA Exoplanet Archive, which is operated by the California Institute of Technology, under contract with the National Aeronautics and Space Administration under the Exoplanet Exploration Program. The research shown here acknowledges use of the Hypatia Catalog Database, an online compilation of stellar abundance data as described in Hinkel et al. (2014, AJ, 148, 54), which was supported by NASA's Nexus for Exoplanet System Science (NExSS) research coordination network and the Vanderbilt Initiative in Data-Intensive Astrophysics (VIDA).

\end{acknowledgments}


\vspace{5mm}
\facilities{Keck:II(NIRSPEC/KPIC)}


\software{astropy \citep{astropy:2013, astropy:2018},  
          \dynesty\ \citep{speagle2020},
          \texttt{corner} \citep{corner},
          \petit\ \citep{prt:2019, prt:2020}}


\appendix
\section{Corner Plots}\label{app:corner}

\begin{figure}
    \centering
    \includegraphics[width=1.0\linewidth]{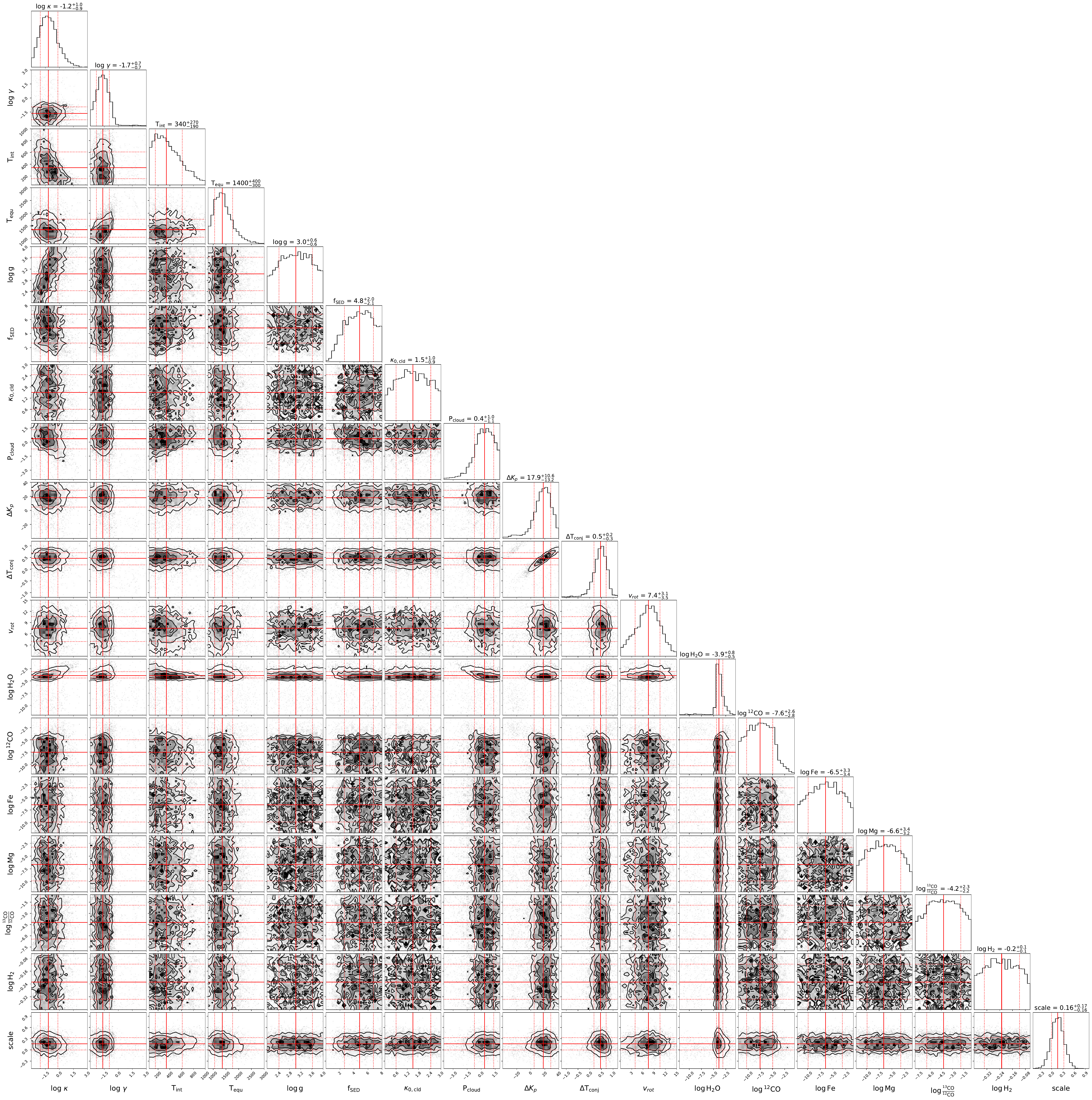}
    \caption{Full corner plot for the 6-component retrieval.}
    \label{fig:corner}
\end{figure}

Figure \ref{fig:corner} presents the full corner plot from the retrieval. 

\clearpage
\bibliography{exoplanetbib}{}

\begin{thebibliography}{}
\expandafter\ifx\csname natexlab\endcsname\relax\def\natexlab#1{#1}\fi
\providecommand{\url}[1]{\href{#1}{#1}}
\providecommand{\dodoi}[1]{doi:~\href{http://doi.org/#1}{\nolinkurl{#1}}}
\providecommand{\doeprint}[1]{\href{http://ascl.net/#1}{\nolinkurl{http://ascl.net/#1}}}
\providecommand{\doarXiv}[1]{\href{https://arxiv.org/abs/#1}{\nolinkurl{https://arxiv.org/abs/#1}}}

\bibitem[{{Astropy Collaboration} {et~al.}(2013){Astropy Collaboration}, {Robitaille}, {Tollerud}, {Greenfield}, {Droettboom}, {Bray}, {Aldcroft}, {Davis}, {Ginsburg}, {Price-Whelan}, {Kerzendorf}, {Conley}, {Crighton}, {Barbary}, {Muna}, {Ferguson}, {Grollier}, {Parikh}, {Nair}, {Unther}, {Deil}, {Woillez}, {Conseil}, {Kramer}, {Turner}, {Singer}, {Fox}, {Weaver}, {Zabalza}, {Edwards}, {Azalee Bostroem}, {Burke}, {Casey}, {Crawford}, {Dencheva}, {Ely}, {Jenness}, {Labrie}, {Lim}, {Pierfederici}, {Pontzen}, {Ptak}, {Refsdal}, {Servillat}, \& {Streicher}}]{astropy:2013}
{Astropy Collaboration}, {Robitaille}, T.~P., {Tollerud}, E.~J., {et~al.} 2013, \aap, 558, A33, \dodoi{10.1051/0004-6361/201322068}

\bibitem[{{Astropy Collaboration} {et~al.}(2018){Astropy Collaboration}, {Price-Whelan}, {Sip{\H{o}}cz}, {G{\"u}nther}, {Lim}, {Crawford}, {Conseil}, {Shupe}, {Craig}, {Dencheva}, {Ginsburg}, {Vand erPlas}, {Bradley}, {P{\'e}rez-Su{\'a}rez}, {de Val-Borro}, {Aldcroft}, {Cruz}, {Robitaille}, {Tollerud}, {Ardelean}, {Babej}, {Bach}, {Bachetti}, {Bakanov}, {Bamford}, {Barentsen}, {Barmby}, {Baumbach}, {Berry}, {Biscani}, {Boquien}, {Bostroem}, {Bouma}, {Brammer}, {Bray}, {Breytenbach}, {Buddelmeijer}, {Burke}, {Calderone}, {Cano Rodr{\'\i}guez}, {Cara}, {Cardoso}, {Cheedella}, {Copin}, {Corrales}, {Crichton}, {D'Avella}, {Deil}, {Depagne}, {Dietrich}, {Donath}, {Droettboom}, {Earl}, {Erben}, {Fabbro}, {Ferreira}, {Finethy}, {Fox}, {Garrison}, {Gibbons}, {Goldstein}, {Gommers}, {Greco}, {Greenfield}, {Groener}, {Grollier}, {Hagen}, {Hirst}, {Homeier}, {Horton}, {Hosseinzadeh}, {Hu}, {Hunkeler}, {Ivezi{\'c}}, {Jain}, {Jenness}, {Kanarek}, {Kendrew}, {Kern}, {Kerzendorf}, {Khvalko}, {King}, {Kirkby}, {Kulkarni},
  {Kumar}, {Lee}, {Lenz}, {Littlefair}, {Ma}, {Macleod}, {Mastropietro}, {McCully}, {Montagnac}, {Morris}, {Mueller}, {Mumford}, {Muna}, {Murphy}, {Nelson}, {Nguyen}, {Ninan}, {N{\"o}the}, {Ogaz}, {Oh}, {Parejko}, {Parley}, {Pascual}, {Patil}, {Patil}, {Plunkett}, {Prochaska}, {Rastogi}, {Reddy Janga}, {Sabater}, {Sakurikar}, {Seifert}, {Sherbert}, {Sherwood-Taylor}, {Shih}, {Sick}, {Silbiger}, {Singanamalla}, {Singer}, {Sladen}, {Sooley}, {Sornarajah}, {Streicher}, {Teuben}, {Thomas}, {Tremblay}, {Turner}, {Terr{\'o}n}, {van Kerkwijk}, {de la Vega}, {Watkins}, {Weaver}, {Whitmore}, {Woillez}, {Zabalza}, \& {Astropy Contributors}}]{astropy:2018}
{Astropy Collaboration}, {Price-Whelan}, A.~M., {Sip{\H{o}}cz}, B.~M., {et~al.} 2018, \aj, 156, 123, \dodoi{10.3847/1538-3881/aabc4f}

\bibitem[{{August} {et~al.}(2023){August}, {Bean}, {Zhang}, {Lunine}, {Xue}, {Line}, \& {Smith}}]{august2023}
{August}, P.~C., {Bean}, J.~L., {Zhang}, M., {et~al.} 2023, \apjl, 953, L24, \dodoi{10.3847/2041-8213/ace828}

\bibitem[{{Baxter} {et~al.}(2020){Baxter}, {D{\'e}sert}, {Parmentier}, {Line}, {Fortney}, {Arcangeli}, {Bean}, {Todorov}, \& {Mansfield}}]{baxter2020}
{Baxter}, C., {D{\'e}sert}, J.-M., {Parmentier}, V., {et~al.} 2020, \aap, 639, A36, \dodoi{10.1051/0004-6361/201937394}

\bibitem[{{Bean} {et~al.}(2023){Bean}, {Xue}, {August}, {Lunine}, {Zhang}, {Thorngren}, {Tsai}, {Stassun}, {Schlawin}, {Ahrer}, {Ih}, \& {Mansfield}}]{bean2023}
{Bean}, J.~L., {Xue}, Q., {August}, P.~C., {et~al.} 2023, \nat, 618, 43, \dodoi{10.1038/s41586-023-05984-y}

\bibitem[{{Birkby} {et~al.}(2017){Birkby}, {de Kok}, {Brogi}, {Schwarz}, \& {Snellen}}]{birkby2017}
{Birkby}, J.~L., {de Kok}, R.~J., {Brogi}, M., {Schwarz}, H., \& {Snellen}, I.~A.~G. 2017, \aj, 153, 138, \dodoi{10.3847/1538-3881/aa5c87}

\bibitem[{{Booth} {et~al.}(2017){Booth}, {Clarke}, {Madhusudhan}, \& {Ilee}}]{booth2017}
{Booth}, R.~A., {Clarke}, C.~J., {Madhusudhan}, N., \& {Ilee}, J.~D. 2017, \mnras, 469, 3994, \dodoi{10.1093/mnras/stx1103}

\bibitem[{{Brogi} {et~al.}(2014){Brogi}, {de Kok}, {Birkby}, {Schwarz}, \& {Snellen}}]{brogi2014}
{Brogi}, M., {de Kok}, R.~J., {Birkby}, J.~L., {Schwarz}, H., \& {Snellen}, I.~A.~G. 2014, \aap, 565, A124, \dodoi{10.1051/0004-6361/201423537}

\bibitem[{{Brogi} \& {Line}(2019)}]{brogi2019}
{Brogi}, M., \& {Line}, M.~R. 2019, \aj, 157, 114, \dodoi{10.3847/1538-3881/aaffd3}

\bibitem[{{Brogi} {et~al.}(2012){Brogi}, {Snellen}, {de Kok}, {Albrecht}, {Birkby}, \& {de Mooij}}]{brogi2012}
{Brogi}, M., {Snellen}, I. A.~G., {de Kok}, R.~J., {et~al.} 2012, \nat, 486, 502, \dodoi{10.1038/nature11161}

\bibitem[{{Brogi} {et~al.}(2013){Brogi}, {Snellen}, {de Kok}, {Albrecht}, {Birkby}, \& {de Mooij}}]{brogi2013}
{Brogi}, M., {Snellen}, I.~A.~G., {de Kok}, R.~J., {et~al.} 2013, \apj, 767, 27, \dodoi{10.1088/0004-637X/767/1/27}

\bibitem[{{Brogi} {et~al.}(2023){Brogi}, {Emeka-Okafor}, {Line}, {Gandhi}, {Pino}, {Kempton}, {Rauscher}, {Parmentier}, {Bean}, {Mace}, {Cowan}, {Shkolnik}, {Wardenier}, {Mansfield}, {Welbanks}, {Smith}, {Fortney}, {Birkby}, {Zalesky}, {Dang}, {Patience}, \& {D{\'e}sert}}]{brogi2023}
{Brogi}, M., {Emeka-Okafor}, V., {Line}, M.~R., {et~al.} 2023, \aj, 165, 91, \dodoi{10.3847/1538-3881/acaf5c}

\bibitem[{{Buchner} {et~al.}(2014){Buchner}, {Georgakakis}, {Nandra}, {Hsu}, {Rangel}, {Brightman}, {Merloni}, {Salvato}, {Donley}, \& {Kocevski}}]{buchner2014}
{Buchner}, J., {Georgakakis}, A., {Nandra}, K., {et~al.} 2014, \aap, 564, A125, \dodoi{10.1051/0004-6361/201322971}

\bibitem[{{Buzard} {et~al.}(2020){Buzard}, {Finnerty}, {Piskorz}, {Pelletier}, {Benneke}, {Bender}, {Lockwood}, {Wallack}, {Wilkins}, \& {Blake}}]{buzard2020}
{Buzard}, C., {Finnerty}, L., {Piskorz}, D., {et~al.} 2020, \aj, 160, 1, \dodoi{10.3847/1538-3881/ab8f9c}

\bibitem[{{Chachan} {et~al.}(2023){Chachan}, {Knutson}, {Lothringer}, \& {Blake}}]{chachan2023}
{Chachan}, Y., {Knutson}, H.~A., {Lothringer}, J., \& {Blake}, G.~A. 2023, \apj, 943, 112, \dodoi{10.3847/1538-4357/aca614}

\bibitem[{{Cridland} {et~al.}(2019){Cridland}, {van Dishoeck}, {Alessi}, \& {Pudritz}}]{cridland2019}
{Cridland}, A.~J., {van Dishoeck}, E.~F., {Alessi}, M., \& {Pudritz}, R.~E. 2019, \aap, 632, A63, \dodoi{10.1051/0004-6361/201936105}

\bibitem[{{Cutri} {et~al.}(2003){Cutri}, {Skrutskie}, {van Dyk}, {Beichman}, {Carpenter}, {Chester}, {Cambresy}, {Evans}, {Fowler}, {Gizis}, {Howard}, {Huchra}, {Jarrett}, {Kopan}, {Kirkpatrick}, {Light}, {Marsh}, {McCallon}, {Schneider}, {Stiening}, {Sykes}, {Weinberg}, {Wheaton}, {Wheelock}, \& {Zacarias}}]{cutri2003}
{Cutri}, R.~M., {Skrutskie}, M.~F., {van Dyk}, S., {et~al.} 2003, VizieR Online Data Catalog, II/246

\bibitem[{{Delorme} {et~al.}(2021){Delorme}, {Jovanovic}, {Echeverri}, {Mawet}, {Kent Wallace}, {Bartos}, {Cetre}, {Wizinowich}, {Ragland}, {Lilley}, {Wetherell}, {Doppmann}, {Wang}, {Morris}, {Ruffio}, {Martin}, {Fitzgerald}, {Ruane}, {Schofield}, {Suominen}, {Calvin}, {Wang}, {Magnone}, {Johnson}, {Sohn}, {L{\'o}pez}, {Bond}, {Pezzato}, {Sayson}, {Chun}, \& {Skemer}}]{kpic}
{Delorme}, J.-R., {Jovanovic}, N., {Echeverri}, D., {et~al.} 2021, Journal of Astronomical Telescopes, Instruments, and Systems, 7, 035006, \dodoi{10.1117/1.JATIS.7.3.035006}

\bibitem[{Echeverri {et~al.}(2022)Echeverri, Jovanovic, Delorme, Xin, Schofield, Finnerty, Wang, Xuan, Mawet, Baker, Bartos, Bond, Bryan, Calvin, Cetre, Doppmann, Fitzgerald, Fucik, Horstman, Lopez, Martin, Martin, Mennesson, Morris, Nash, Pezzato, Porter, Ragland, Roberts, Ruane, Ruffio, Sappey, Serabyn, Skemer, Venenciano, Wallace, Wang, \& Wizinowich}]{echeverri2022}
Echeverri, D., Jovanovic, N., Delorme, J.-R., {et~al.} 2022, in Ground-based and Airborne Instrumentation for Astronomy IX, ed. C.~J. Evans, J.~J. Bryant, \& K.~Motohara, Vol. 12184, International Society for Optics and Photonics (SPIE), 121841W, \dodoi{10.1117/12.2630518}

\bibitem[{{Espinoza} {et~al.}(2017){Espinoza}, {Fortney}, {Miguel}, {Thorngren}, \& {Murray-Clay}}]{espinoza2017}
{Espinoza}, N., {Fortney}, J.~J., {Miguel}, Y., {Thorngren}, D., \& {Murray-Clay}, R. 2017, \apjl, 838, L9, \dodoi{10.3847/2041-8213/aa65ca}

\bibitem[{{ExoFOP}(2019)}]{exofop3}
{ExoFOP}. 2019, Exoplanet Follow-up Observing Program - TESS,  IPAC, \dodoi{10.26134/EXOFOP3}

\bibitem[{{Feroz} \& {Hobson}(2008)}]{feroz2008}
{Feroz}, F., \& {Hobson}, M.~P. 2008, \mnras, 384, 449, \dodoi{10.1111/j.1365-2966.2007.12353.x}

\bibitem[{{Feroz} {et~al.}(2009){Feroz}, {Hobson}, \& {Bridges}}]{feroz2009}
{Feroz}, F., {Hobson}, M.~P., \& {Bridges}, M. 2009, \mnras, 398, 1601, \dodoi{10.1111/j.1365-2966.2009.14548.x}

\bibitem[{{Feroz} {et~al.}(2019){Feroz}, {Hobson}, {Cameron}, \& {Pettitt}}]{feroz2019}
{Feroz}, F., {Hobson}, M.~P., {Cameron}, E., \& {Pettitt}, A.~N. 2019, The Open Journal of Astrophysics, 2, 10, \dodoi{10.21105/astro.1306.2144}

\bibitem[{Finnerty {et~al.}(2022)Finnerty, Schofield, Delorme, Sappey, Wang, Ruffio, Mawet, Fitzgerald, Jovanovic, Baker, Bartos, Bond, Bryan, Calvin, Cetre, Doppmann, Echeverri, Lopez, Martin, Morris, Pezzato, Ragland, Ruane, Skemer, Venenciano, Wallace, Wang, Wizinowich, \& Xuan}]{Finnerty2022}
Finnerty, L., Schofield, T., Delorme, J.-R., {et~al.} 2022, in Ground-based and Airborne Instrumentation for Astronomy IX, ed. C.~J. Evans, J.~J. Bryant, \& K.~Motohara, Vol. 12184, International Society for Optics and Photonics (SPIE), 121844Y, \dodoi{10.1117/12.2630276}

\bibitem[{{Finnerty} {et~al.}(2023){Finnerty}, {Schofield}, {Sappey}, {Xuan}, {Ruffio}, {Wang}, {Delorme}, {Blake}, {Buzard}, {Fitzgerald}, {Baker}, {Bartos}, {Bond}, {Calvin}, {Cetre}, {Doppmann}, {Echeverri}, {Jovanovic}, {Liberman}, {L{\'o}pez}, {Martin}, {Mawet}, {Morris}, {Pezzato}, {Phillips}, {Ragland}, {Skemer}, {Venenciano}, {Wallace}, {Wallack}, {Wang}, \& {Wizinowich}}]{finnerty2023}
{Finnerty}, L., {Schofield}, T., {Sappey}, B., {et~al.} 2023, \aj, 166, 31, \dodoi{10.3847/1538-3881/acda91}

\bibitem[{{Finnerty} {et~al.}(2024){Finnerty}, {Xuan}, {Xin}, {Liberman}, {Schofield}, {Fitzgerald}, {Agrawal}, {Baker}, {Bartos}, {Blake}, {Calvin}, {Cetre}, {Delorme}, {Doppmann}, {Echeverri}, {Hsu}, {Jovanovic}, {L{\'o}pez}, {Martin}, {Mawet}, {Morris}, {Pezzato}, {Ruffio}, {Sappey}, {Skemer}, {Venenciano}, {Wallace}, {Wallack}, {Wang}, \& {Wang}}]{finnerty2024}
{Finnerty}, L., {Xuan}, J.~W., {Xin}, Y., {et~al.} 2024, \aj, 167, 43, \dodoi{10.3847/1538-3881/ad1180}

\bibitem[{Foreman-Mackey(2016)}]{corner}
Foreman-Mackey, D. 2016, The Journal of Open Source Software, 1, 24, \dodoi{10.21105/joss.00024}

\bibitem[{{Fortney} {et~al.}(2008){Fortney}, {Lodders}, {Marley}, \& {Freedman}}]{fortney2008}
{Fortney}, J.~J., {Lodders}, K., {Marley}, M.~S., \& {Freedman}, R.~S. 2008, \apj, 678, 1419, \dodoi{10.1086/528370}

\bibitem[{{Fu} {et~al.}(2024){Fu}, {Welbanks}, {Deming}, {Inglis}, {Zhang}, {Lothringer}, {Ih}, {Moses}, {Schlawin}, {Knutson}, {Henry}, {Greene}, {Sing}, {Savel}, {Kempton}, {Louie}, {Line}, \& {Nixon}}]{fu2024}
{Fu}, G., {Welbanks}, L., {Deming}, D., {et~al.} 2024, arXiv e-prints, arXiv:2407.06163, \dodoi{10.48550/arXiv.2407.06163}

\bibitem[{{Gaia Collaboration}(2020)}]{gaiaedr3}
{Gaia Collaboration}. 2020, VizieR Online Data Catalog, I/350

\bibitem[{{Gandhi} {et~al.}(2024){Gandhi}, {Landman}, {Snellen}, {Welbanks}, {Madhusudhan}, \& {Brogi}}]{gandhi2024}
{Gandhi}, S., {Landman}, R., {Snellen}, I., {et~al.} 2024, \mnras, 530, 2885, \dodoi{10.1093/mnras/stae1048}

\bibitem[{{Gandhi} {et~al.}(2023){Gandhi}, {Kesseli}, {Zhang}, {Louca}, {Snellen}, {Brogi}, {Miguel}, {Casasayas-Barris}, {Pelletier}, {Landman}, {Maguire}, \& {Gibson}}]{gandhi2023}
{Gandhi}, S., {Kesseli}, A., {Zhang}, Y., {et~al.} 2023, \aj, 165, 242, \dodoi{10.3847/1538-3881/accd65}

\bibitem[{{Gordon} {et~al.}(2022){Gordon}, {Rothman}, {Hargreaves}, {Hashemi}, {Karlovets}, {Skinner}, {Conway}, {Hill}, {Kochanov}, {Tan}, {Wcis{\l}o}, {Finenko}, {Nelson}, {Bernath}, {Birk}, {Boudon}, {Campargue}, {Chance}, {Coustenis}, {Drouin}, {Flaud}, {Gamache}, {Hodges}, {Jacquemart}, {Mlawer}, {Nikitin}, {Perevalov}, {Rotger}, {Tennyson}, {Toon}, {Tran}, {Tyuterev}, {Adkins}, {Baker}, {Barbe}, {Can{\`e}}, {Cs{\'a}sz{\'a}r}, {Dudaryonok}, {Egorov}, {Fleisher}, {Fleurbaey}, {Foltynowicz}, {Furtenbacher}, {Harrison}, {Hartmann}, {Horneman}, {Huang}, {Karman}, {Karns}, {Kassi}, {Kleiner}, {Kofman}, {Kwabia-Tchana}, {Lavrentieva}, {Lee}, {Long}, {Lukashevskaya}, {Lyulin}, {Makhnev}, {Matt}, {Massie}, {Melosso}, {Mikhailenko}, {Mondelain}, {M{\"u}ller}, {Naumenko}, {Perrin}, {Polyansky}, {Raddaoui}, {Raston}, {Reed}, {Rey}, {Richard}, {T{\'o}bi{\'a}s}, {Sadiek}, {Schwenke}, {Starikova}, {Sung}, {Tamassia}, {Tashkun}, {Vander Auwera}, {Vasilenko}, {Vigasin}, {Villanueva}, {Vispoel}, {Wagner}, {Yachmenev}, \&
  {Yurchenko}}]{hitemp2020}
{Gordon}, I.~E., {Rothman}, L.~S., {Hargreaves}, R.~J., {et~al.} 2022, \jqsrt, 277, 107949, \dodoi{10.1016/j.jqsrt.2021.107949}

\bibitem[{{Guillot}(2010)}]{guillot2010}
{Guillot}, T. 2010, \aap, 520, A27, \dodoi{10.1051/0004-6361/200913396}

\bibitem[{{Guilluy} {et~al.}(2019){Guilluy}, {Sozzetti}, {Brogi}, {Bonomo}, {Giacobbe}, {Claudi}, \& {Benatti}}]{guilluy2019}
{Guilluy}, G., {Sozzetti}, A., {Brogi}, M., {et~al.} 2019, \aap, 625, A107, \dodoi{10.1051/0004-6361/201834615}

\bibitem[{{H{\'e}brard} {et~al.}(2016){H{\'e}brard}, {Arnold}, {Forveille}, {Correia}, {Laskar}, {Bonfils}, {Boisse}, {D{\'\i}az}, {Hagelberg}, {Sahlmann}, {Santos}, {Astudillo-Defru}, {Borgniet}, {Bouchy}, {Bourrier}, {Courcol}, {Delfosse}, {Deleuil}, {Demangeon}, {Ehrenreich}, {Gregorio}, {Jovanovic}, {Labrevoir}, {Lagrange}, {Lovis}, {Lozi}, {Moutou}, {Montagnier}, {Pepe}, {Rey}, {Santerne}, {S{\'e}gransan}, {Udry}, {Vanhuysse}, {Vigan}, \& {Wilson}}]{hebrard2016}
{H{\'e}brard}, G., {Arnold}, L., {Forveille}, T., {et~al.} 2016, \aap, 588, A145, \dodoi{10.1051/0004-6361/201527585}

\bibitem[{{Hinkel} {et~al.}(2014){Hinkel}, {Timmes}, {Young}, {Pagano}, \& {Turnbull}}]{hypatia}
{Hinkel}, N.~R., {Timmes}, F.~X., {Young}, P.~A., {Pagano}, M.~D., \& {Turnbull}, M.~C. 2014, \aj, 148, 54, \dodoi{10.1088/0004-6256/148/3/54}

\bibitem[{{Hubeny} {et~al.}(2003){Hubeny}, {Burrows}, \& {Sudarsky}}]{hubeny2003}
{Hubeny}, I., {Burrows}, A., \& {Sudarsky}, D. 2003, \apj, 594, 1011, \dodoi{10.1086/377080}

\bibitem[{{Kempton} \& {Knutson}(2024)}]{kempton2024}
{Kempton}, E. M.~R., \& {Knutson}, H.~A. 2024, Reviews in Mineralogy and Geochemistry, 90, 411, \dodoi{10.2138/rmg.2024.90.12}

\bibitem[{{Khorshid} {et~al.}(2022){Khorshid}, {Min}, {D{\'e}sert}, {Woitke}, \& {Dominik}}]{khorshid2022}
{Khorshid}, N., {Min}, M., {D{\'e}sert}, J.~M., {Woitke}, P., \& {Dominik}, C. 2022, \aap, 667, A147, \dodoi{10.1051/0004-6361/202141455}

\bibitem[{{Lei} \& {Molli{\`e}re}(2024)}]{lei2024}
{Lei}, E., \& {Molli{\`e}re}, P. 2024, arXiv e-prints, arXiv:2410.21364, \dodoi{10.48550/arXiv.2410.21364}

\bibitem[{{Li} {et~al.}(2015){Li}, {Gordon}, {Rothman}, {Tan}, {Hu}, {Kassi}, {Campargue}, \& {Medvedev}}]{li2015}
{Li}, G., {Gordon}, I.~E., {Rothman}, L.~S., {et~al.} 2015, \apjs, 216, 15, \dodoi{10.1088/0067-0049/216/1/15}

\bibitem[{{Ligterink} {et~al.}(2024){Ligterink}, {Kipfer}, \& {Gavino}}]{ligterink}
{Ligterink}, N.~F.~W., {Kipfer}, K.~A., \& {Gavino}, S. 2024, \aap, 687, A224, \dodoi{10.1051/0004-6361/202450405}

\bibitem[{{Line} {et~al.}(2021){Line}, {Brogi}, {Bean}, {Gandhi}, {Zalesky}, {Parmentier}, {Smith}, {Mace}, {Mansfield}, {Kempton}, {Fortney}, {Shkolnik}, {Patience}, {Rauscher}, {D{\'e}sert}, \& {Wardenier}}]{line2021}
{Line}, M.~R., {Brogi}, M., {Bean}, J.~L., {et~al.} 2021, \nat, 598, 580, \dodoi{10.1038/s41586-021-03912-6}

\bibitem[{{Lodders}(2003)}]{lodder2003}
{Lodders}, K. 2003, \apj, 591, 1220, \dodoi{10.1086/375492}

\bibitem[{{L{\'o}pez} {et~al.}(2020){L{\'o}pez}, {Hoffman}, {Doppmann}, {Fitzgerald}, {Johnson}, {Kassis}, {Lanclos}, {Lyke}, {Martin}, {McLean}, {Sohn}, \& {Weiss}}]{nirspecupgrade2}
{L{\'o}pez}, R.~A., {Hoffman}, E.~B., {Doppmann}, G., {et~al.} 2020, in Society of Photo-Optical Instrumentation Engineers (SPIE) Conference Series, Vol. 11447, Society of Photo-Optical Instrumentation Engineers (SPIE) Conference Series, 114476B, \dodoi{10.1117/12.2563075}

\bibitem[{{Lothringer} {et~al.}(2018){Lothringer}, {Barman}, \& {Koskinen}}]{lothringer2018}
{Lothringer}, J.~D., {Barman}, T., \& {Koskinen}, T. 2018, \apj, 866, 27, \dodoi{10.3847/1538-4357/aadd9e}

\bibitem[{{Lothringer} {et~al.}(2021){Lothringer}, {Rustamkulov}, {Sing}, {Gibson}, {Wilson}, \& {Schlaufman}}]{lothringer2021}
{Lothringer}, J.~D., {Rustamkulov}, Z., {Sing}, D.~K., {et~al.} 2021, \apj, 914, 12, \dodoi{10.3847/1538-4357/abf8a9}

\bibitem[{{Malsky} {et~al.}(2021){Malsky}, {Rauscher}, {Kempton}, {Roman}, {Long}, \& {Harada}}]{malsky2021}
{Malsky}, I., {Rauscher}, E., {Kempton}, E. M.~R., {et~al.} 2021, \apj, 923, 62, \dodoi{10.3847/1538-4357/ac2a2a}

\bibitem[{{Martin} {et~al.}(2018){Martin}, {Fitzgerald}, {McLean}, {Doppmann}, {Kassis}, {Aliado}, {Canfield}, {Johnson}, {Kress}, {Lanclos}, {Magnone}, {Sohn}, {Wang}, \& {Weiss}}]{nirspecupgrade}
{Martin}, E.~C., {Fitzgerald}, M.~P., {McLean}, I.~S., {et~al.} 2018, in Society of Photo-Optical Instrumentation Engineers (SPIE) Conference Series, Vol. 10702, Ground-based and Airborne Instrumentation for Astronomy VII, ed. C.~J. {Evans}, L.~{Simard}, \& H.~{Takami}, 107020A, \dodoi{10.1117/12.2312266}

\bibitem[{Mawet {et~al.}(2019)Mawet, Fitzgerald, Konopacky, Beichman, Jovanovic, Dekany, Hover, Chisholm, Ciardi, Artigau, Banyal, Beatty, Benneke, Blake, Burgasser, Canalizo, Chen, Do, Doppmann, Doyon, Dressing, Fang, Greene, Hillenbrand, Howard, Kane, Kataria, Kempton, Knutson, Kotani, Lafreniere, Liu, Nishiyama, Pandey, Plavchan, Prato, Rajaguru, Robertson, Salyk, Sato, Schlawin, Sengupta, Sivarani, Skidmore, Tamura, Terada, Vasisht, Wang, \& Zhang}]{hispec}
Mawet, D., Fitzgerald, M., Konopacky, Q., {et~al.} 2019, arXiv e-prints, \dodoi{10.48550/ARXIV.1908.03623}

\bibitem[{{McLean} {et~al.}(1998){McLean}, {Becklin}, {Bendiksen}, {Brims}, {Canfield}, {Figer}, {Graham}, {Hare}, {Lacayanga}, {Larkin}, {Larson}, {Levenson}, {Magnone}, {Teplitz}, \& {Wong}}]{nirspec}
{McLean}, I.~S., {Becklin}, E.~E., {Bendiksen}, O., {et~al.} 1998, in Society of Photo-Optical Instrumentation Engineers (SPIE) Conference Series, Vol. 3354, Infrared Astronomical Instrumentation, ed. A.~M. {Fowler}, 566--578, \dodoi{10.1117/12.317283}

\bibitem[{{Molli{\`e}re} {et~al.}(2017){Molli{\`e}re}, {van Boekel}, {Bouwman}, {Henning}, {Lagage}, \& {Min}}]{molliere2017}
{Molli{\`e}re}, P., {van Boekel}, R., {Bouwman}, J., {et~al.} 2017, \aap, 600, A10, \dodoi{10.1051/0004-6361/201629800}

\bibitem[{{Molli{\`e}re} {et~al.}(2019){Molli{\`e}re}, {Wardenier}, {van Boekel}, {Henning}, {Molaverdikhani}, \& {Snellen}}]{prt:2019}
{Molli{\`e}re}, P., {Wardenier}, J.~P., {van Boekel}, R., {et~al.} 2019, \aap, 627, A67, \dodoi{10.1051/0004-6361/201935470}

\bibitem[{{Molli{\`e}re} {et~al.}(2020){Molli{\`e}re}, {Stolker}, {Lacour}, {Otten}, {Shangguan}, {Charnay}, {Molyarova}, {Nowak}, {Henning}, {Marleau}, {Semenov}, {van Dishoeck}, {Eisenhauer}, {Garcia}, {Garcia Lopez}, {Girard}, {Greenbaum}, {Hinkley}, {Kervella}, {Kreidberg}, {Maire}, {Nasedkin}, {Pueyo}, {Snellen}, {Vigan}, {Wang}, {de Zeeuw}, \& {Zurlo}}]{prt:2020}
{Molli{\`e}re}, P., {Stolker}, T., {Lacour}, S., {et~al.} 2020, \aap, 640, A131, \dodoi{10.1051/0004-6361/202038325}

\bibitem[{Nasedkin {et~al.}(2024)Nasedkin, Mollière, \& Blain}]{Nasedkin2024}
Nasedkin, E., Mollière, P., \& Blain, D. 2024, Journal of Open Source Software, 9, 5875, \dodoi{10.21105/joss.05875}

\bibitem[{{{\"O}berg} {et~al.}(2011){{\"O}berg}, {Murray-Clay}, \& {Bergin}}]{oberg2011}
{{\"O}berg}, K.~I., {Murray-Clay}, R., \& {Bergin}, E.~A. 2011, \apjl, 743, L16, \dodoi{10.1088/2041-8205/743/1/L16}

\bibitem[{{Pai Asnodkar} {et~al.}(2022){Pai Asnodkar}, {Wang}, {Eastman}, {Cauley}, {Gaudi}, {Ilyin}, \& {Strassmeier}}]{pai2022}
{Pai Asnodkar}, A., {Wang}, J., {Eastman}, J.~D., {et~al.} 2022, \aj, 163, 155, \dodoi{10.3847/1538-3881/ac51d2}

\bibitem[{{Pelletier} {et~al.}(2021){Pelletier}, {Benneke}, {Darveau-Bernier}, {Boucher}, {Cook}, {Piaulet}, {Coulombe}, {Artigau}, {Lafreni{\`e}re}, {Delisle}, {Allart}, {Doyon}, {Donati}, {Fouqu{\'e}}, {Moutou}, {Cadieux}, {Delfosse}, {H{\'e}brard}, {Martins}, {Martioli}, \& {Vandal}}]{pelletier2021}
{Pelletier}, S., {Benneke}, B., {Darveau-Bernier}, A., {et~al.} 2021, \aj, 162, 73, \dodoi{10.3847/1538-3881/ac0428}

\bibitem[{{Pelletier} {et~al.}(2023){Pelletier}, {Benneke}, {Ali-Dib}, {Prinoth}, {Kasper}, {Seifahrt}, {Bean}, {Debras}, {Klein}, {Bazinet}, {Hoeijmakers}, {Kesseli}, {Lim}, {Carmona}, {Pino}, {Casasayas-Barris}, {Hood}, \& {St{\"u}rmer}}]{pelletier2023}
{Pelletier}, S., {Benneke}, B., {Ali-Dib}, M., {et~al.} 2023, \nat, 619, 491, \dodoi{10.1038/s41586-023-06134-0}

\bibitem[{{Polyansky} {et~al.}(2018){Polyansky}, {Kyuberis}, {Zobov}, {Tennyson}, {Yurchenko}, \& {Lodi}}]{polyansky2018}
{Polyansky}, O.~L., {Kyuberis}, A.~A., {Zobov}, N.~F., {et~al.} 2018, \mnras, 480, 2597, \dodoi{10.1093/mnras/sty1877}

\bibitem[{{Ramkumar} {et~al.}(2023){Ramkumar}, {Gibson}, {Nugroho}, {Maguire}, \& {Fortune}}]{ramkumar2023}
{Ramkumar}, S., {Gibson}, N.~P., {Nugroho}, S.~K., {Maguire}, C., \& {Fortune}, M. 2023, \mnras, 525, 2985, \dodoi{10.1093/mnras/stad2476}

\bibitem[{{Rodler} {et~al.}(2012){Rodler}, {Lopez-Morales}, \& {Ribas}}]{rodler2012}
{Rodler}, F., {Lopez-Morales}, M., \& {Ribas}, I. 2012, \apjl, 753, L25, \dodoi{10.1088/2041-8205/753/1/L25}

\bibitem[{{Roth} {et~al.}(2024){Roth}, {Parmentier}, \& {Hammond}}]{roth2024}
{Roth}, A., {Parmentier}, V., \& {Hammond}, M. 2024, \mnras, 531, 1056, \dodoi{10.1093/mnras/stae984}

\bibitem[{{Rothman} {et~al.}(2010){Rothman}, {Gordon}, {Barber}, {Dothe}, {Gamache}, {Goldman}, {Perevalov}, {Tashkun}, \& {Tennyson}}]{hitemp2010}
{Rothman}, L.~S., {Gordon}, I.~E., {Barber}, R.~J., {et~al.} 2010, \jqsrt, 111, 2139, \dodoi{10.1016/j.jqsrt.2010.05.001}

\bibitem[{{Speagle}(2020)}]{speagle2020}
{Speagle}, J.~S. 2020, \mnras, 493, 3132, \dodoi{10.1093/mnras/staa278}

\bibitem[{{Spiegel} {et~al.}(2009){Spiegel}, {Silverio}, \& {Burrows}}]{spiegel2009}
{Spiegel}, D.~S., {Silverio}, K., \& {Burrows}, A. 2009, \apj, 699, 1487, \dodoi{10.1088/0004-637X/699/2/1487}

\bibitem[{{Webb} {et~al.}(2020){Webb}, {Brogi}, {Gandhi}, {Line}, {Birkby}, {Chubb}, {Snellen}, \& {Yurchenko}}]{webb2020}
{Webb}, R.~K., {Brogi}, M., {Gandhi}, S., {et~al.} 2020, \mnras, 494, 108, \dodoi{10.1093/mnras/staa715}

\bibitem[{{Weiner Mansfield} {et~al.}(2024){Weiner Mansfield}, {Line}, {Wardenier}, {Brogi}, {Bean}, {Beltz}, {Smith}, {Zalesky}, {Batalha}, {Kempton}, {Montet}, {Owen}, {Plavchan}, \& {Rauscher}}]{weinermansfield2024}
{Weiner Mansfield}, M., {Line}, M.~R., {Wardenier}, J.~P., {et~al.} 2024, \aj, 168, 14, \dodoi{10.3847/1538-3881/ad4a5f}

\bibitem[{{Wiser} {et~al.}(2024){Wiser}, {Line}, {Welbanks}, {Mansfield}, {Parmentier}, {Bean}, \& {Fortney}}]{wiser2024}
{Wiser}, L.~S., {Line}, M.~R., {Welbanks}, L., {et~al.} 2024, \apj, 971, 33, \dodoi{10.3847/1538-4357/ad5097}

\end{thebibliography}
\bibliographystyle{aasjournal}



\end{CJK*}
\end{document}